\documentstyle[11pt,epsf,aaspp4]{article}

\tighten
\slugcomment{}

\begin{document}

\def\la{\lower.5ex\hbox{$\; \buildrel < \over \sim \;$}}
\def\ga{\lower.5ex\hbox{$\; \buildrel > \over \sim \;$}}
\def\r{\hangindent=1pc \noindent}
\def\j21{J_{912,-21}}
\def\lyal {{\rm Ly} \alpha}
\def\lya {{\rm Ly} \alpha}
\def\be {\begin{equation}}
\def\bee {\begin{equation*}}
\def\ee {\end{equation}}
\def\eee {\end{equation*}}
\def\omigm {\Omega_{\scriptscriptstyle\rm IGM}}
\def \nh{N _{\scriptscriptstyle\rm H \,I}}
\def \nhe{N _{\scriptscriptstyle\rm He \,II}}
\def \whmin{W^{\scriptscriptstyle\rm H \, I} _{
\scriptscriptstyle\rm min}}
\def \ang{{\rm \AA}}
\def\h{H~I}
\def\he{He~I}
\def\hei{He \,II}

\def\etal{{et~al.}}
\def\eg{{e.g.~}}
\def\ie{{i.e.~}}
\def\noi{\noindent}
\def\bs{\bigskip}
\def\ms{\medskip}
\def\ss{\smallskip}
\def\ob{\obeylines}
\def\l{\line}
\def\hrf{\hrulefill}
\def\hf{\hfil}
\def\q{\quad}
\def\qq{\qquad}
\renewcommand{\deg}{$^{\circ}$}
\newcommand{\um}{$\mu$m}
\newcommand{\uk}{$\mu$K}
\newcommand{\qrms}{$Q_{rms-PS}$}
\newcommand{\n}{$n$}
\newcommand{\cdmr}{${\bf c}_{\rm DMR}$}
\newcommand{\xrms}{$\otimes_{RMS}$}
\newcommand{\gt}{$>$}
\newcommand{\lt}{$<$}
\newcommand{\ldl}{$< \delta <$}


\title{Photoelectric heating for dust grains at high redshifts}
 
\author{Biman B. Nath$^1$, 
 Shiv K. Sethi$^2$, Yuri Shchekinov$^{3,4}$}
\bs
\affil{$^1${\it Raman Research Institute, Bangalore --560080,
India}}
\affil{$^2${\it Institut d'Astrophysique, 98 bis, Boulevard Arago, 75014
Paris, France}}
\affil{$^3${\it Osservatorio Astrofisico di Arcetri, Largo E. Fermi, 5,
50125 Firenze, Italy}}
\affil{$^4${\it Department of Physics, Rostov State University, Sorge 5, 
344090, Rostov on Don, Russia}}
\bs
\affil{(biman@rri.ernet.in, 
sethi@iap.fr, yus@rsuss1.rnd.runnet.ru)}

\begin{abstract}
Lyman-$\alpha$ absorption systems at $z \sim 3$ with 
$\nh \ga 3 \times 10^{14}$
cm$^{-2}$ have been found to be enriched  with a mean metallicity of
$Z/Z_{\odot} \sim 10^{-2.5}$, and a large scatter in the metallicity. 
It is reasonable to assume that the process of
initial enrichment of the intergalactic
medium (IGM) at $z \ga 3$ also produced dust grains. We explore the
implications of the presence of dust grains in the IGM at high redshift,
in particular, the contribution of photoelectric emission from grains 
by hard background photons to
the net heating rate of the IGM. 

We show that (i) the charge on dust particles and the characteristics
of photo-emitted electrons differ substantially from those in the interstellar
medium (ISM) 
in several respects: a) grains are exposed to and charged by photons 
beyond the Lyman limit, and b) because of this, the photoelectrons have 
typical energy of tens of eV. We also show that
ii) silicates are more efficient heating agents than graphites; iii) small 
grains contribute mostly to the net heating; iv) at densities typical of 
the IGM at $z\sim 3$ and for $\lya$ absorbers, dust heating can be 
comparable or exceed photoionization heating within an order of
magnitude; v) this increases the temperature of overdense 
regions, compared to the case of no dust heating, 
by factor of $\sim 2$. 
We discuss the implications of this extra heating source in $\lyal$
absorbing systems.
\end{abstract}


\section{Introduction}

Recent observations with high resolution have 
provided evidences for the existence of heavy elements in $\lya$ forest
clouds at redshifts $z\sim 3$ (Cowie et al., 1995, Tytler et al., 
1995). Songaila
and Cowie (1996) claimed that more than $75\%$ of all $\lyal$ absorbers
with $\nh \ga 10^{14}$ cm$^{-2}$  at $z \sim 3$ are enriched. Tytler
et al. (1995) reported similar results : all $\lya$ systems with $\nh \ga
10^{15}$ cm$^{-2}$ and 60\% with $\nh \ga 10^{14}$ cm$^{-2}$. These
values are expected to be lower limits, since 
the measurements of heavy elements are S/N limited, 
and all $\lya$ systems may contain heavy elements (see also Tytler 1988).

The abundance of metals, however, remains somewhat uncertain, especially
because it depends on the value of the ionization parameter ($\Gamma$, the
ratio of ionizing photon density to the particle density) assumed. 
Nevertheless,
for a reasonable range of $\Gamma=10^{-1.5}\hbox{--}10^{-2.5}$, the inferred
metallicity is almost independent of the H~I column density, and is of
order $Z/Z_{\odot} \sim 10^{-2}$ (Songaila \& Cowie 1996). Rauch et al. (1997)
found from the comparison of the data with their numerical simulations that
the metallicity has a mean $\sim 10^{-2.5}$ solar and that there is a large
scatter. Songaila (1997) has recently reported a lower limit to the metallicity 
of order $\Omega_{metals}/\Omega_b \sim 1\hbox{--}4 \times 10^{-5} \, h \, 
(1+2q_o z)^{1/2}$. For a low $q_o$ universe ($q_o=0.02$), $h=0.65$ and a solar
metallicity of $0.019$, this implies a minimum metallicity of order
$10^{-3.3}\hbox{--}10^{-2.8}$ solar. 

Recent studies of metallicity and dust content in nearby galaxies have 
demonstrated that dust-to-gas ratio $k$ and metallicity $\zeta=Z/Z_\odot$ 
correlate 
(Schmidt \& Boller, 1993, Lisenfeld \& Ferrara, 1997). Though the 
correlation is not always good, it definitely shows that the production of
metals in galaxies is accompanied with dust formation. Moreover, weak 
correlation between $k$ and $\zeta$ in blue compact dwarf galaxies may  
indicate that galaxies during active phases of star formation eject dust 
along with metals in gaseous outflows (Lisenfeld \& Ferrara, 1997).  
It is, therefore, reasonable to expect that if the IGM and $\lyal$ absorbers
are enriched with metals, they are also contaminated 
with dust grains, produced inside
the galaxies and transported into the IGM along with metals. 
The presence of dust in the IGM and its 
consequences was first discussed by Rowan-Robinson et al. (1979), 
who suggested that dust in the early universe at $z\sim 200$ could lead 
to distortions of CMBR (cosmic microwave background radiation)
(see also Vainer 1990; Loeb \& Haiman 1997). 
More recently, Fall et al. (1989) and 
Pei et al. (1991) have
argued that the reddening of background quasars indicates directly that
dust grains exist in damped $\lya$ systems. Their dust-to-gas ratio 
ranges from $k=0.06$ to $k=0.35$ for different
extinction curves assumed (here $k=10^{21}\tau_B/\nh$, $k_G=0.79$ for the
Galaxy, Fall \& Pei, 1989). Arguments based on the extinction of light by 
dust have led Zou et al. (1997) to similar estimates of $k$ in damped 
$\lya$ clouds toward the lensed quasar 0957+561. They explain the difference
in the continuum shape of the two spectra by differential reddening due
to the presence of dust grains. Recently, however, Pettini et al. (1997) have 
shown that the dust-to-metals ratio in damped $\lya$ systems is
only $\simeq 1/2$ of its value in local interstellar clouds (thus, 
dust-to-gas ratio is $k\simeq k_G \zeta/2$). They infer 
this from the fact that Cr and other refractory elements are
depleted only by a factor of $2$ in 18 studied damped $\lya$ systems, much less 
than in the interstellar medium, and assuming that this depletion is due 
to freezing on dust grains. 

In this paper we assume 
that dust particles are present not only in damped 
$\lya$ systems, but also in $\lya$ forest clouds, following the idea
that dust is associated with metals. We concentrate here on detailed
study of dust charging in the high redshift IGM illuminated by the
background UV and X-ray photons and heating due to injected
supra-thermal photo-electrons---in several respects these processes 
differ from those in the local interstellar medium. 

We do not discuss here how metals and dust particles are supplied 
to the IGM; we refer the reader to the papers where the IGM enrichment 
has been 
associated with
population III stars (Ciardi \& Ferrara, 1997, Gnedin \& Ostriker 1997),
early-forming very massive objects (Carr, Bond \& Arnett 1984),  
galactic winds from galaxies at high redshifts (Silk, Wyse \& Shields 1987;
Miralda-Escud\'e \& Rees 1997; Nath \& Trentham 1997).
The estimates show
that an enriched IGM with a metallicity of $10^{-2.5}$ Z$_{\odot}$ at
$z \sim 3$ can be understood in terms of either Population III objects or
galactic winds.

The paper is structured as follows.
In \S 2 we describe the basic
processes and the equations, in \S 3 we describe the basic characteristics 
of charged high-redshift dust particles exposed to the background 
UV radiation, and photo-emitted electrons,  and in \S 4 we present the results
of the evolution of temperature in the IGM with photo-induced dust
heating as an important energy source. We then
briefly discuss the implication of our results for $\lyal$ absorbers.
Throughout the paper we use a Hubble constant $H_0=h_{50} \, 50$ km s$^{-1}$
Mpc$^{-1}$ and $\Omega=1$.

\section{The charge of a dust grain in the IGM}

\subsection{The UV background radiation}

We first discuss the ambient photon background that is important for the
process of charging of the dust at high redshift. The existence of a UV
background radiation at high redshift has been inferred from proximity effect
observations (Bajtlik et al. 1988). These observations show that the UV
background radiation has an intensity at 
$\lambda=912 \> {\rm \AA}$ of $\j21 \sim 0.3 
\hbox{--} 1$, in the units of $10^{-21}$ erg s$^{-1}$ cm$^{-2}$ sr$^{-1}$ 
Hz$^{-1}$ ($2 < z< 3$) (Bechtold 1994; Cristiani et al. 1995; Cooke et al. 1996).

\begin{table}
\centerline{Table 1 : Models for the UV background radiation}
\vskip 0.2in
\begin{center}
\begin{tabular}{|c||c||c|}
\hline
$\lambda ({\rm \AA})$ & QSO & QSO+YG \\
\hline
\hline
$\lambda =912 ({\rm \AA})$ & $J_{-21}=0.5$ & $J_{-21}=1.$\\
\hline
$228 < \lambda ({\rm \AA}) <912$ & $J_{\nu} \propto \nu^{-0.5}$ &
                                 $J_{\nu} \propto \nu^{-2.5}$\\
\hline
$\lambda =228 ({\rm \AA})$ & $J_{-21}=0.025$ & $J_{-21}=0.00123$\\
\hline
$\lambda < 228({\rm \AA})$ & $J_{\nu} \propto \nu^{-1}$ &
                                 $J_{\nu} \propto \nu^{-1}$\\
\hline
$S_L(\equiv J_{912}/J_{228})$ & $20.$ & $813.$\\
\hline
\end{tabular}
\end{center}
\end{table}

On the theoretical front, following the pioneering work of 
Miralda-Escud\'e \&
Ostriker (1990), several authors have calculated the expected value of 
$\j21$ from the quasars,
using existing quasar luminosity functions, and taking into account the 
absorption of photons by the intervening $\lyal$ absorption systems (e.g.,
Madau 1992; Haardt and Madau 1996). This predicted intensity, however, 
falls short of
the inferred  intensity from the proximity effect (Cooke, Espey and Carswell
1996). It has, therefore, been suggested that there are other sources of
UV photons at high redshift, {\it e.g.} young star forming galaxies, decaying
neutrinos, or even a warm, collisionally ionized IGM. 
The predicted {\it spectrum}
of the UV background in these cases however is much different, and, therefore,
observational constraints on the spectrum hold the clue to the origin of 
the high redshift UV background radiation. 
The observations of He~II absorption at high redshift put constraints on 
the spectrum of
the ionizing background (Jakobsen \etal 1994; Davidsen \etal 1996). Meiksin \&
Madau (1994) showed that the He~II opacity observed by Jakobsen \etal (1994)
softness ratio,  $S_L \equiv J_{912}/J_{228} \ga 40$ at $z \sim 3.3$ assuming
a reasonable value of the opacity due to diffuse H~I 
(see also, Songaila, Hu \& Cowie 1995).
For the Davidsen \etal (1996) observations, Sethi and Nath (1997) 
found that the allowed range of 
models is still large, after taking various uncertainties into account
($16 \la S_L \la 650$).
There has been recent claims that the spectral 
index changes around $z \sim 3$
(Songaila 1998, Reimers \etal 1997, Hogan \etal 1997, Savaglio \etal 1997),
but we assume a constant spectral index
for simplicity. We discuss the effect of such a change in the section
\S 3.3.

For our purpose here, we will use two models for the UV background 
radiation for our calculation, bearing in mind the above uncertainties. 
The first
one (referred to as QSO) represents the background radiation 
expected from only
quasars, and the second one (QSO+YG), represents the radiation 
expected from quasars
and young galaxies.
The models are described in Table 1. The model QSO is a broken power-law
fit to the spectrum
obtained by Haardt and Madau (1996), for an intermediate
redshift of $z \sim 2.5$. Note that their calculation was actually
for $q_o=0.1$. The model QSO+YG assumes the QSO luminosity function
of Boyle (1991), as in Madau (1992), and  
a population of young galaxies, which is used to make $J_{912,-21}=1$ 
using the A3 model
of absorption of Giroux \& Shapiro (1996) (again at $z \sim 2.5$; 
see also Sethi and Nath 1997). The values of the softness ratio $S_L$ for 
each model is given in Table 1. Although the broken power laws are
crude approximations to the actual spectrum, they suffice for our
purpose of illustrating the effect of dust charging and heating at
high redshift, in the light of the uncertainty in our knowledge of
the UV background radiation.

Beyond $228 \> {\rm \AA}$, we use $J_\nu \propto \nu^{-1}$ till $h\nu=3$ keV
for both models. This again mimics well the Haardt \& Madau (1996) spectra for
$10 \, {\rm \AA} \la \lambda \la 228 \, {\rm \AA}$.
For $E_\gamma \ga 3$ keV, we use the quasar X-ray
spectral index of $0.9$. Zdziarski \etal (1995) found a 
good fit to the data of Seyferts with a spectral index of $0.9$ till 
$\sim 100$ keV. However, photons with energies much above $1$ keV do not
contribute significantly in the charging and heating processes due to
small absorption coefficients at these energies.

The evolution of the background radiation in redshift also
remains uncertain. For
simplicity, we assume that $J_{912,-21}$
remains a constant for $2<z<5$, and $J_{912,-21} \propto
(1+z)^{2.8}$ for $z<2$.
This is a fit to the evolution of $J_{912,-21}$ as calculated
by Madau (1992) for $z<2$.
Also, we assume that the spectral shape remains a constant
in time, although in reality the spectrum becomes softer at higher 
redshift due to increase in absorption (Miralda-Escud\'e \& 
\& Ostriker 1990). Again, however, the uncertainties involved are
large, and we assume a constant spectral shape for simplicity.

\subsection{Basic processes for charging grains} 

In this paper we follow Draine (1978) and Draine \& Salpeter (1979, 
hereafter DS) in 
describing the processes determining the grain charge. The equation 
governing the charge on a grain is written as follows 

\be
{dq\over dt}=4\pi a^2e\Sigma_i J_i\>,  
\ee
where $J_i$ is the partial current due to particle of the $i$-th kind, and
$a$ is the grain radius. The currents which mainly determine grain charge in 
the ISM are due to: {\it i)} impinging electrons, {\it ii)} 
impinging ions (mainly protons), {\it iii)}  UV photons. In 
principle, secondary processes can be associated with these three (such 
as secondary electron emission), however, in the conditions we are 
interested in they play a minor role. In case of incident electrons and 
protons, currents $J_i$ are enhanced due to Coulomb interaction 
by factor $g(x)$. For electrons, with $x=eU/kT$, where $U$ is the grain
potential,
(Spitzer, 1978), 

\be
g_e(x)=\left\{ \begin{array}{ll}
                \exp(x) & \mbox{if $x<0$} \\
                1+x  &  \mbox{if $x>0 \> .$}
               \end{array}
       \right.
\ee
For ions, it is given by,

\be
g_i(x)=\left\{ \begin{array}{ll}
                \exp(-z x) & \mbox{if $x>0$} \\
                1-z x  &  \mbox{if $x<0 \> ,$}
               \end{array}
       \right.
\ee
where $z$ is the ion charge.

\subsubsection{The current due to impinging electrons}

Grain particles in low density IGM exposed by UV and X-ray photons 
are positively charged, and the electron current can be written as 
(DS), 

\be J_e=-n_e\left({kT\over 2\pi m_e}\right)^{1/2}
(\langle s_e \rangle -\langle \delta \rangle)(1+x) \>,
\ee
here $\langle s_e \rangle$, the thermally averaged sticking 
probability for incident electrons. This probability,
 $\langle s_e \rangle \simeq 1$ for 
thermal electrons with temperature $T$ and grain size $a$ satisfying
the condition (Draine, 1989), 

\be a> 10\left({T\over 10^6~{\rm K}}\right)^{1.5}~\AA \>, 
\ee 
which is valid for temperature $T\leq 10^5$ K, the range of temperatures
relevant in our case. $\langle\delta\rangle$ is the secondary emission
coefficient, for incident thermal electrons with $kT\ll 300-500$ 
eV $ \langle \delta \rangle \ll 1$. 

\subsubsection{The current due to impinging ions (protons)}

DS give, for the proton current, 

\be J_{ip}=n_p\left({kT\over 2\pi M}\right)^{1/2}
(z \langle s_p \rangle+ \langle \nu_p \rangle -\beta 
\langle Y_p \rangle )\exp(-x) \>. 
\ee
where $\langle s_p \rangle \simeq 1$, the effective sticking coefficient,  
$\langle \nu_p \rangle\ll 1$, the secondary electron emission coefficient,
and $\beta \langle Y_z \rangle\ll 1$, the 
charging due to sputtered charge particles, as can be estimated from DS
for temperature $T\leq 10^5$ K.
It is seen clearly that in fully ionized IGM, contribution of protons 
to grain charging is of order $|J_p| \sim 0.03 |J_e|$. We neglect the current 
from helium ions, since for the situations we are concerned with in this 
paper, the contribution from ions is in general small compared to
electron and photoinduced currents. 

\subsubsection{Photoelectric current per area} 

Photoelectric current per grain area is 

\be J_{UV}=\int\limits_{\nu_{min}}^\infty Q(a,\nu) y_\nu {4 \pi J_\nu 
\over h\nu} d\nu \>, 
\ee 
where $J_\nu$ is the spectral energy flux of incident photons 
([erg cm$^{-2}$ s$^{-1}$ Hz$^{-1}$ sr$^{-1}$]), 
$y_\nu$, the photo-yield, {\it i.e.} the number of photo-emitted electrons per 
absorbed photon, $Q_\nu(a)$, the efficiency of photoabsorptions, and
$\nu_{min} = (w + eU)/h$ corresponds to the minimum energy for which electrons
can escape from the surface of a dust grain, $w$ being the work function. 
The normalized photo-yield can be written 
as (Draine, 1978, Bakes \& Tielens, 1994, hereafter BT) ,

\be y_\nu=y_0\left(1-{I_z\over h\nu}\right)f_z(a) \>, 
\ee
where (Draine, 1978),

\be
f_z(a)=\left(\zeta\over \alpha\right)^2 
{(\alpha^2-2 \alpha+2 -2 e^{-\alpha})\over 
(\zeta^2-2\zeta +2 -2 e^{-\zeta})} \>, 
\ee
with 

\be \alpha={a\over l_a}+{a\over l_e};\qquad \zeta={a\over l_a} \>, 
\ee
where $a$ is the grain radius, $l_a$, the photon attenuation length, 
$l_e$, the electron escape length; BT give $l_a=10^{-6}$ cm, and 
$l_e=10^{-7}$ cm. The asymptotic (at photon energies much larger than 
the ionization potential of a grain $I_z$) value of 
$y_0$ is determined by Draine (1978) as $y_0=0.5$ to reproduce the 
enhanced value of the photo-yield of a small particle over the 
bulk graphite, and by BT as $y_0=0.14$ to fit experimental data for
coronene (Verstraete et al, 1990). 
$I_z$ is the ionization potential for a grain with charge 
$Z$ (BT), and is given by,

\be
I_z=w+(Z+{1\over 2}){e^2\over a} \>. 
\ee
Here  $e$ is in esu units. 
It is important to note here 
that when the photo-induced current is calculated for the ISM in the
Galaxy, the upper limit in the integral is defined as the Lyman limit 
frequency $\nu_L$ (Drain, 1978, BT), regarding the fact 
that Lyman continuum photons 
are confined within HII regions. This restricts from above the charge of
dust grains, the energy of photo-emitted electrons and thus the
photoelectric heating. In our case, charging of dust grains
and photoelectric heating are determined by photons with energy above 
Lyman limit. 
BT stress that near threshold the photo-yield for graphite grains 
is usually overestimated. In our case, 
with photon energies and electric potentials much higher than 
in the ISM, the contribution from the threshold region is small.

\subsubsection{Photoabsorption efficiency} 

For silicate particles $Q$ can be reasonably approximated  
for sizes what we are concerning $a\leq 0.1~ \mu{\rm m}$, 
and photon energies $\epsilon \geq 10$ eV by (Ferrara \& 
Dettmar, 1994) 

\be Q(a,\epsilon)=\left({\epsilon\over 8~{\rm eV}}\right)^{\gamma_a}
\>, 
\ee
where 

\be \gamma_a=2.5 \Bigl[(10a)^{0.1}-1 
\Bigr]\>,
\ee
here and throughout the paper $a$ is in $\mu$m, if not specified. 

For graphite particles, we use the approximation, 

\begin{eqnarray}
Q(a,\epsilon)&=&30\times a
\exp(-11.9\sqrt{a})\sigma_1(\epsilon) \nonumber\\&+&
60\times 
{a\over 1+25 a} 
\exp(-7.5\sqrt{a})\sigma_2(\epsilon)\nonumber\\&+&
{10 a\over (1+a)(1+10a)}
\sigma_3(\epsilon)
, 
\end{eqnarray}
where $\sigma_i$, the cross sections per carbon atom in a particle 
(in MegaBarns), are described below.
For $\sigma_1$ and $\sigma_2$, we used the data 
given in BT (see their Fig. 1), which we approximated as 

\be
\sigma_1(\epsilon)=\cases{ 
5.24\times 10^{-4}\epsilon^{3.78}, \qq  
\epsilon \leq 17.29 \, {\rm eV} \cr
1.37\times 10^{5}\epsilon^{-3.02}, \qq 
\epsilon > 17.29 \,  {\rm eV}}\>;  
\ee

\be
\sigma_2(\epsilon)=\cases{0.032\epsilon^{2.5},
\qq \epsilon<6.13 \, {\rm eV}\cr 
112.7\epsilon^{-2}
,\qq \epsilon> 6.13 \,{\rm eV}}\>;
\ee
\be
\sigma_3(\epsilon)=1.5+2\times 10^{-4}(\epsilon-0.22)
(\epsilon-18)(\epsilon-40)\>,
\ee
here $\epsilon$ is given in eV.
This approximation for $Q(a,\epsilon)$ agrees   with  the 
results of Draine and Lee (1984) (see their Fig.~4) for grain sizes
 $a=0.03-0.1$ and with  the results of BT (see their Fig.~1),
 which are based on an  extrapolation of 
the results of Draine and  Lee to  smaller particles,  
to within 10--15 \%, for
$3 \, {\rm eV} \la \epsilon \la 40 \, {\rm eV}$, and 20--30 \% for 
$\epsilon \la 3 \, \rm eV$. BT showed that the ionization cross sections
obtained in this way are in good agreement with the data for small PAH
coronene (Verstraete et al. 1990).

At energies $ \epsilon> 40 \, \rm  eV$, for $Q(a,\epsilon)$, we use 
the approximation given by DS  
\be
Q(a,\epsilon)=Q(a, 40{\rm ~eV})\left({40~{\rm eV}\over
\epsilon}\right)\>;
\ee
 and for $\epsilon>1$ keV, 
\be
Q(a,\epsilon)=Q(a,1~{\rm keV})\left({1~{\rm keV}\over 
\epsilon}\right)^3\>.
\ee
These approximations give reasonable fits (within 10--15 \%) in the entire
energy range.  One should stress however, that in the 
high-energy interval, 
$\epsilon \ga 1$ keV, the $Q(a,\epsilon)\propto \epsilon^{-3}$ dependence 
is contaminated by very sharp increases in $Q$ when the photon
energy approaches the edge of an atomic subshell, and can exceed 
the $Q(a,\epsilon)\propto \epsilon^{-3}$ value by 1--1.5 orders of magnitude,
(Dwek \& Smith, 1996). From this point of view our results for dust 
charging and heating can be considered conservative. 

\subsection{Grain potential} 

It is readily seen from the charge equation that characteristic 
charging times caused by the photo-induced and electron currents, 
for $n_e \ge 10^{-7}$ cm$^{-3}$, are 
much shorter than the Hubble time at $z \la 20$ ($t_H\sim 10^{15}$ s)
and sound crossing time for a cloud with $R>1$ kpc with temperature 
$T\sim 10^4-10^5$ K ($t_R\sim 10^{15}$ s). 
Therefore, the grain charge 
is kept at equilibrium, $\dot q=0$, and can be easily determined from 

\be
n_e \Bigl ( {8kT \over \pi m_e} \Bigr ) ^{1/2} 
(1+x)= \int^{\infty} _{\nu _{min}} Q(a,\nu) y_{\nu}
{4 \pi J_{\nu} \over h \nu} d \nu +
n_p \Bigl ( {8kT \over \pi m_p} \Bigr ) ^{1/2}\>.
\ee
It is clear that the proton current can be comparable to the
photo-induced current when the particle flux for protons $\sim n_pv_p$ is 
comparable with the photon flux $\sim J_\nu/h$. More precisely, 
for a power law spectrum $J_\nu\propto \nu^{-\alpha}$ it is 
equivalent to the condition, 

\be
\left({8kT\over \pi m_p}\right)^{1/2}n_p\sim
{4\pi QyJ_0\over \alpha h}\left({h\nu_L\over w}\right)^\alpha
\>, 
\ee
where $J_0$ is the spectral energy flux at the Lyman limit
frequency, $\nu_L$, 
$Q$ and $y$, the absorption efficiency and 
photo-yield averaged over the spectrum. 
For typical values $J_0\sim 10^{-21}$, $Q\sim 0.1$, $y\sim 0.1$ this 
is equivalent to $n_e\sim 0.2 T_4^{-1/2}$. Thus, for conditions in the IGM
at $z\sim 3$ photo-induced current dominates dust charging. 

\subsubsection{Analytical estimates}

For grains exposed to hard UV radiation, the charge 
and potential are expected to be high, $x\gg 1$, and can be
estimated for a power-law spectrum 
of incident photons $J\propto \nu^{-\alpha}$ and a power-law dependence
for $Q(a,\nu)\propto Q(a,\nu_L)(\nu/\nu_L)^{-\beta}$, as 

\be
x\sim \Biggl[{ 10^{\alpha+\beta-2}\over 
n_eT_4^{\alpha+\beta+1/2} (\alpha+\beta)}Q(a,\nu_L)\Biggr]^{1/(\alpha+\beta+1)} \>.
\ee
Here, we have used $y=0.1$; our assumption of $x\gg 1$ is valid for
$n_e < 10^{-3}-10^{-4}$ cm$^{-3}$, and
$T_4< 10$. For $\alpha
\sim 1$, $\beta\sim 3$ (such $\beta$ is typical for graphites
particles of small sizes in the energy interval $\epsilon=17.3-40$ eV),
this gives a weak dependence of the potential on gas density and
absorption characteristics, 
$x\sim 2 n_e^{-0.2}T_4^{-0.9}[J_{-21}Q(a,\nu_L)]^{0.2}$.
In $\lya$ clouds with $n_e\sim 10^{-4}$ cm$^{-3}$ and $T\sim 10^4$ K
$x\sim 10 [J_{-21}Q(a,\nu_L)]^{0.2}$ which is much larger than 
the value for grains
in the local ISM illuminated by photons below Lyman limit.
For graphite particles with larger radius, $a\sim 0.1 $,
$\beta\sim 1-1.5$, and $x\sim 0.8n_e^{-0.25}T_4^{-0.6}
[J_{-21}Q(a,\nu_L)]^{0.25}-$
$n_e^{-0.28}T_4^{-0.86}[J_{-21}Q(a,\nu_L)]^{0.28}$. 

Silicate particles have
less strong dependence of $Q(a,\epsilon)$ on photon energy in the interval
$\epsilon> 10$ eV, $\beta\sim 0- 1$ for $a=0.1-0.001$, respectively, 
which results in a stronger dependence of grain potential on
the parameters of ambient medium: $x \sim 0.3n_e^{-0.5}T_4^{-0.75}
[J_{-21}Q(a,\nu_L)]^{0.5}$--
$0.8n_e^{-0.33}T_4^{-0.83}[J_{-21}Q(a,\nu_L)]^{0.33}$.
Note that for the dependence of the graphites grain potential
 on $Q(a)$ and on grain size is weaker than for silicates.

\subsubsection{Numerical results}

Figures 1a and 1b show the dependence of grain potential $x=eU/kT$
for graphites and silicates, respectively, on the ambient density
for hard (QSO, solid lines) and soft (QSO+YG, dashed lines) spectra
of background photons, for a grain size of $a=10^{-5}$ cm
and for fixed gas temperature $T=10^4$ K. Figures
2a and 2b show the dependence of $x$ for graphites and silicates
on the grain size $a$ for different ambient densities for the QSO
and QSO+YG spectra.

The figures show that the dependences follow qualitatively the above
simple estimates. In general, hard spectrum (QSO) results in higher
value of $x$, especially for silicates, and both graphites and silicates
have comparatively lower grain potentials for soft spectrum. QSO 
spectra with low equivalent $\alpha$ result in more strong dependence 
of grain potential on density than QSO+YG spectra with larger equivalent
$\alpha$ do. Graphites
do not show much dependence of the potential on the grain size, while
silicates do, which is due to more weak dependence of $Q(a,\epsilon)$ 
on photon energy for silicates, as seen from analytical estimates. 
It is also seen that grain potential on silicate particles depends on 
the ambient density stronger than on graphite particles. 


%

\subsection{Photo-electric heating} 

The heating rate due to photo-electric effect in dust grains 
(in erg cm$^{-3}$ s$^{-1}$) is given by,

\begin{equation}
H_{ph}=n_H \int da {\cal N}_g(a)
\sigma(a) \int\limits_{\nu_{m1(a)}}^\infty d\nu Q(a,\nu) y_\nu 
{4 \pi J_\nu \over h \nu}
(h \nu -w-eU(a)) \>,
\label{eq:heat}
\end{equation}
where 
$\nu_{m1(a)}=(w+eU+3kT/2)/h$, the minimum frequency of 
photons needed to
eject electrons which have sufficient energy to heat the surrounding gas, 
${\cal N}_g(a)=dn/da$, the size distribution of grains, is taken  
a power-law ${\cal N}_g(a)=Aa^{-q}$.
The normalizing 
factor $A$ can be determined from 

\be A\int\limits_{a_{min}}^{a_{max}} a^{-q}{4\over 3}\pi 
\rho_g a^3 da=10^{-2}\zeta m_Hn \>, 
\ee
where $a_{min}=5\times 10^{-7}$ cm, $a_{max}=4\times 10^{-5}$ cm 
(see, O'Donnel \& Mathis, 1997), are the minimal and maximal sizes 
of dust particles, respectively. Then, the coefficient $A$ is 


\be A=3.8\times 10^{-27}(4-q)\left({Z\over Z_\odot}\right) 
{n\over a_{max}^{4-q}-a_{min}^{4-q}}\>,
\label{eq:norm}
\ee
for typical grain density $\rho_g=1$ g cm$^{-3}$.  
For $U(a)$ in Eq.~\ref{eq:heat},
we use the results for grain charge as obtained in \S 2.3.

\subsubsection{Mean energy of photoelectrons}

An important parameter which determines the photo-electric heating is 
the mean energy of photo-emitted electrons, 

\begin{equation}
E(a)= {\int\limits_{\nu_{min}}^\infty d\nu Q(a,\nu) y_\nu 
{4 \pi J_\nu  \over  h \nu}
(h \nu -w-eU(a))\over  
\int\limits_{\nu_{min}}^\infty d\nu Q(a,\nu) y_\nu 
{4 \pi J_\nu \over  h \nu}} \>.
\end{equation}

\noindent
In the local interstellar medium $E(a)$ is of the order of $h\nu_L-
w \sim 5.6$ eV. In a medium exposed to hard UV photons it is 
restricted only from below, and the exact value depends on the UV 
spectrum. Figures 3 and 4 show this dependence for graphite and silicate
particles. It is seen that $E(a)$ can reach hundreds of eV at low densities,
while typical values for moderate densities is $30-50~kT\sim$ tens
of eV. 

Note that although the curves in Figures 3 \& 4 show that in general
$E(a)$ for grains with higher charges is large, it should not be
taken to mean that
higher charge {\it causes} the mean electron energies to be larger.
The dependence of $E(a)$ on $eU/kT$ at $T=10^4$ K is shown in Figure 5
for silicates (solid lines) and graphites (dotted lines), for QSO type
(thick lines) and QSO+YG type spectra (thin lines) (the upper set of
curves are for $a=10^{-5}$ cm and the lower set of curves are for
$a=10^{-6}$ cm). It is seen that $E(a)$ is a monotonically increasing
function of $eU/kT$, and the nature of the curve depends on $Q (a, \nu)$,
$y_{\nu}$ and the photon spectrum.


The differential heating rate (per unit grain) is a product of 
absorbed photons and 
$E(a)$ 

$$
{dH_{ph}\over dn_g(a)}=
E(a) \int\limits_{\nu_{m1(a)}}^\infty d\nu Q(a,\nu) y_\nu \sigma(a)
{4 \pi J_\nu \over h \nu} \>,
$$
and can be therefore roughly estimated as $dH_{ph}(a)/dn_g(a)
\propto E(a)(eU)^{\beta+\alpha-1}\propto E(a)n_e^{(\beta+\alpha-1)/
(\beta+\alpha+1)}$, \ie flatter than linear.

\subsubsection{Differential heating rates} 

Using the approximate solution for dust potential what we found in
section 2.3.1 one can estimate partial contributions to the 
total heating rate from silicates and graphites. In the integral  

\begin{equation}
{dH_{ph}\over dn_g(a)}=
\int\limits_{\nu_{m1(a)}}^\infty d\nu Q(a,\nu) y_\nu \sigma(a)
{4 \pi J_\nu \over h \nu}
(h \nu -w-eU(a)) \>,
\label{eq:pheat}
\end{equation}
we can neglect $w$ in comparison with $h\nu$ and $eU$, and using 
as above power laws for the flux $J$ and absorption coefficient
$Q(a,\nu)$ we get 

\begin{equation}
{dH_{ph}\over dn_g(a)}\simeq 4\times 10^{-5}f(\alpha,\beta) \sigma(a)
n_e^{\alpha+\beta-1\over \alpha+\beta+1}
T_4^{-{\alpha+\beta-1\over 2(\alpha+\beta+1)}}
[J_{-21}Q(a,\nu_L)]^{2\over \alpha+\beta+1},\>
\label{eq:quahet}
\end{equation}
where $f(\alpha,\beta)\sim 1$. It is seen from this equation that 
the differential heating rate, both for graphite and silicate particles,
depends on grain radius through 
$Q(a,\nu_L)$ rather weakly, 
and thus contribution
from grains of small radius dominates due to the factor ${\cal N}_g(a)
\propto a^{-q}$. However, the integrated heating rate is insensitive to
the grain size distribution: for $Q^{2/\alpha+\beta+1}\sim a^\eta$, 
with $0<\eta<1$ the dependence of heating rate on $q$ is close to be 
eliminated with normalizing factor $A$ in Eq. \ref{eq:norm}. 

In figure 6 we show heating rates (per unit volume) due to 
graphite and silicate particles, respectively, integrated over size 
distributions for both incident
photon spectra, and for different size distribution
of grains, $q=3.5$ and $q=3.9$. For comparison
we show also the photoionization heating rate for a primordial gas.
It is seen that photoelectric
dust heating depends on density less strongly than photoionization heating. 
Silicates produce weaker dependence of heating rate on density, which
can be explained by the weak dependence of grain potential and the
mean energy of photo-electrons on density (see, Figs 2 and 3). Due to 
the fact that both $eU$ and $E(a)$ for silicates are higher than for 
graphites, the heating rate by silicate grains is also higher. Changing
the size distribution results in a weak change (within factor of 1.5) 
of the heating rate: dust with $q=3.9$ heats more efficiently than 
with $q=3.5$, as expected. 
In Figure 7, we show the heating rate as a function of the lower limit
of the grain size assumed, for different grain types and ambient densities.
The curves show that the heating is dominated by small grains in general.
In brief, in all cases the photo-electric 
dust heating 
is comparable or even exceeds (at low densities) the photoionization 
heating. 

The dependence of the dust heating rate on the ionizing radiation flux 
at Lyman edge $J(\nu_L)$ is quite weak, particularly for QSO+YG spectra
with large $\alpha$. However, it can lead to variations in temperature 
due to local inhomogeneity in exposing UV flux. On the other hand, the 
photoionization heating does not depend on the flux in a fully 
ionized medium with $n_{HI}\ll n_H$ for low density ($n \ll 10^{-3.5}$ 
cm$^{-3}$
as considered here), as the photoionization heating rate is proportional
to the recombination rate and thus weakly depends on the UV flux.. 

Dust heating can also stimulate at appropriate conditions spatial 
variations in temperature via thermal instability. In the simplest case of 
steady energy balance: $\Lambda(T)n^2=H_{ph}$, with $H_{ph}$ depending 
on density and temperature as given in Eq.~\ref{eq:quahet}, Field's  
criterion for thermal instability of condensation mode (Field, 1965) 
is written as 

\begin{equation}
{d\log \Lambda\over d\log T}-1+{3\over 2}{\alpha+\beta-1\over
\alpha+\beta+1}<0,\>
\end{equation}
which shows that for 
$\alpha+\beta<1$ (hard spectrum and dominance of silicate grains with 
$a>0.01~\mu$m), dust 
heating works as a destabilizing factor. 
\section{Thermal evolution of the clumpy IGM with dust heating} 

Next, we compute the thermal evolution of the clumps in the IGM with
overdensities of order $\sim 10$, which either keep their overdensity
constant (\ie, adiabatically expand), or are confined by some mechanism
and keep their density constant. These are described
below in detail as Models I and II.

We assume that the gas in $\lyal$ systems evolves passively, driven
by the UV background flux and the expansion (or non-expansion) of the cloud.
This is consistent with the picture of $\lyal$ systems as described
by the recent numerical simulations  (e.g., Miralda-Escud\'e \etal 1996)
in which $\lyal$ system gas is
not involved in star formation and are not affected by the processes
related to it.

We note here that
Miralda-Escud\'e \& Rees (1994) also calculated the evolution of temperature
in overdense regions of IGM, which first collapse under gravity, during
which the temperature rises to $\sim 10^5$ K due to adiabatic heating,
 and which then evolve with
constant (proper) density, during which, due to high density and the
consequent high recombination and line cooling rate, temperature falls down
quickly to $\sim 10^4$ K.They, however, considered much higher densities
($\ga 10^{-3}$ cm$^{-3}$)
than we use here ($\la 10^{-4}$ cm$^{-3}$) (at lower densities, the
balance between the heating and cooling processes will be different, for
example, adiabatic cooling will be more important).

The evolution of temperature in an overdense region in
the IGM would depend on the initial temperature (and epoch) assumed.
If the gas in the overdense region were initially compressed adiabatically,
then the maximum temperature would be $T_{cl}
 \sim (\rho_{cl}/
\rho_{IGM})^{2/3} \, T_{IGM}$, where $\rho_{cl}$ is the density in the
overdense region. For overdensities of order $5\hbox{--}10$, that are
considered to be typical values for $\lyal$ absorbing systems (see below),
$T_{cl} \sim 3\hbox{--}5 \, T_{IGM}$. The value of $T_{IGM}$, however,
depends on the thermal history of the IGM (Miralda-Escud\'e \& Rees 1994)
which remains uncertain. We, therefore, for the sake of illustrating the
importance of dust heating, assume a few values of the initial temperatures
of the overdense region $T_{in}$ which are of order $10^4$ K, and follow
the thermal evolution in time.

\subsection{Models: heating and cooling processes}

In numerical calculations we used the heating rate due to photo-electric 
effect in dust grains as described above with grain size distribution 
${\cal N}_g\propto a^{-q}$ with $q=3.5$. For the relative fraction of 
graphites and silicates, we use the one given by Pei (1992) for the 
Large Magellanic Cloud (LMC), which seems suitable for low metallicity
objects at high redshift (Pei, Fall \& Bechtold 1991). 
We also show the thermal evolutions with only graphites and silicates
involved. The case of only silicates correspond to the Small Magellanic
Cloud (SMC) type dust grains (Pei 1992).
For other heating processes apart from dust heating,
we use photoionization heating  of a primordial gas, for both models
of UV radiation. The heating due to photoionization of carbon atoms
is smaller than this by a factor of $3$ ($10$) for the QSO+YG (QSO) spectrum
for a metallicity of $Z/Z_{\odot}=0.01$.
For cooling, we use recombination cooling,
cooling due to collisional ionization, line cooling (in the presence
of the UV radiation) and bremsstrahlung for a primordial gas, as given
in Cen (1992). Cooling due to heavy elements at this metallicity and
temperature is unimportant. 
Inverse Compton cooling at $z \la 5$ is not important and is neglected here 
(the corresponding cooling time-scale is $t_{iC}
\sim 9 \times 10^{12} (1+z)^{-4}$ yr;
Tegmark, Silk \& Evrard (1993)).

In the Model I,
we assume that the overdense regions adiabatically
expand keeping the overdensity a constant in time. In this case, therefore, we
also include cooling due to adiabatic expansion. In the Model II, we 
consider regions with constant (proper) density, in which the gas is confined 
by gravity or shock ram pressure, or some other confinement mechanism.
For Model I, we use $\rho_{cl}/\Omega_b \rho_c
=10$, for all redshifts, where $\rho_{cl}$ is the
density inside the cloud, $\rho_c$ is the critical density
and $\Omega_b$ is the global mean baryon density in the units of $\rho_c$ .
For Model II, we use $\rho_{cl}/\Omega_b \rho_c=10$  (at $z=z_{in}$) for 
different values of the initial redshift $z_{in}$.
We use $\Omega_b=0.03$ and initial temperature $T_{in}=10^4, 2 \times 10^4,
5 \times 10^4 $ K.
Our choice of the parameters follow the result of Cen \& Simcoe
(1997), that $\lyal$ absorbers with $\nh \sim 10^{14}$
cm$^{-2}$ had $\rho_{cl}/\rho_b \sim 10$ at $z\sim 3$ for 
$\Omega_b \sim 0.03$. The minimum width of $\lyal$ lines at
$z\sim4$ has been recently found to be $b=15$ km s$^{-1}$ (Kim \etal 1997).
If the minimum width (in the distribution of $b$ values) is considered
to be due to gas temperature (see below for more discussion on this),
this corresponds to a temperature of $\sim 10^4$ K and proves the presence
of overdense regions with such temperatures at $z \sim 4$. Here we consider
the temperature evolution of such regions in time.

In reality, the gas in $\lyal$ systems is expected to behave in a way
which lies between our two models.
Numerical simulations show that the structures in the IGM 
corresponding to $\lyal$ forest absorption lines are 
sheet-like structures (Miralda-Escud\'e \etal 1996). The gas 
is confined in one dimension and flows 
along the sheet toward semi-spherical vortices. Therefore, the density
is expected to fall slower than $(1+z)^3$. Also, since these structures
are transient, lasting about a Hubble time, our calculations 
would be relevant only on time scales shorter than this.

\subsection{Models: numerical results}

We show the evolution of the gas temperature in
Figure 8a and 8b for graphites and silicates.
The dotted and solid lines refer to the cases
without dust, and with dust with $\zeta=Z/Z_{\odot}=0.01$. The
plots show the evolution for various initial temperatures, for
the two models of background photons (upper panels for QSO and
lower panels for QSO+YG spectra). The case of silicates, of course,
correspond to the grain mixture of the SMC type (Pei 1992).
In figure 9 we show the case of
grain mixture of the LMC type (Pei 1992) for the QSO spectrum for
three values of initial temperatures and metallicities.

It is seen that the dominant contribution to the heating comes from 
silicate grains for both type of spectra. This is due to the fact that for
silicates, the absorption efficiency does not decrease with increasing
photon energy as sharply as it does for graphites, and therefore, the 
high energy photons contribute substantially in this case. 
In case with hard background
spectrum (QSO), the dust heating is more efficient for the same reasons--- 
due to contribution of high energy photons, and as a result of the high 
mean energy of photo-emitted electrons. In Models I 
without dust heating adiabatic cooling decreases temperature sharply, 
so that at $z=1-2$ it is below 10$^4$ K. Dust heating is able to support
temperature at the level $T\ga 10^4$ down to $z=0$. In Models II with 
dust heating temperature is factor of 1.5--2 higher than without dust 
heating. At $z<1$ both photoionization and photoelectric heating 
decrease due to decrease of the UV flux.

Since dust heating manifests itself better (compared to other sources
of heating) at lower densities, it is conceivable that the diffuse,
low density, `intercloud' component of the IGM, if any, would have
a much different thermal history with dust than with only photoionization
heating. The density of such a component remains, however, uncertain.
Recent numerical studies show that most of the baryons in the IGM
are in the overdense regions, leaving about $5 \%$ in the diffuse
component (Zhang \etal 1997). Their results show a fragmented IGM
at high redshift, and they claim that even the `diffuse' medium is also
possibly fragmented into scales smaller than the resolution of the 
simulation. For $\Omega_b=0.03$ and $h_{50}=1$, this corresponds to
a density  $ \simeq 4 \times 10^{-9} (1+z)^3$ cm$^{-3}$ (using
$5\%$ of the baryons in the diffuse component). Judging from the
heating curves of Fig. 6, such a component would be hotter than
if only photoionization is used. However, the temperature
would depend on the thermal history of the IGM.

\subsection{Uncertainties}

Here we discuss briefly the effect of uncertainties in the parameters
on our results. Since the type of dust particles in $\lyal$ systems
is uncertain, we showed the temperature evolution for different types
of dust (graphite and silicate) in Figures 8a and 8b. It is seen that
the effect on the temperature of the gas is by factors of order $\la 1.5$.
We also show the effect of different metallicities on the temperature
evolution in Figure 9, for $Z/Z_{\odot}=0.003, 0.01$. The difference
in the temperature of the gas is again of order $\sim 1.5$. Since the
actual temperature evolution of the gas would depend on the initial
conditions, the readers are urged to use the heating rate shown in
Figure 6, and scale it for the required metallicity for different
cases (the heating rate is proportional to the metallicity).

Finally, the evolution of the UV background radiation is still uncertain.
For example, it has been claimed that the spectrum changes at $z \sim 3$
from soft at higher redshifts to a harder spectrum later (Songaila 1998 and
the references therein).
We have already shown the heating rates (Figure 6) and temperature
evolution of the gas (Figures 8a and 8b) for a hard (QSO) and a soft
(QSO+YG) spectra, with the spectral index being constant in time.
Figures 8a and 8b show that the temperature of the gas at later redshifts
($z \sim 2$) differ by very small factors for different values of $z_{in}$.
Since the heating rate for the soft spectrum case is very small compared
to the hard spectrum case, a change in the spectrum at $z \sim 3$ would
essentially be equivalent to the case of $z _{in}\sim 3$,
and as explained above, it would effect the temperature evolution at later
redshifts only
negligibly. For $z_{in}=4$, the temperature would evolve much like the
case of just photoionization between $3 \la z \la 4$
(dotted lines in Figures 8a and 8b).

Similarly, the possibility that the value of $\j21$ drops by a factor
of $3\hbox{--}10$ from $z \sim 2.5$ to $z \sim 4.2$ (Williger \etal 1994,
Kim \etal 1997), would essentially mean a very small heating rate at $z \ga 3$,
but as explained above, it would not effect the temperature evolution
at later redshifts much.

\section{Discussions}

Finally, we discuss various implications of our results on photoelectric
heating due to dust in IGM. To begin with, we discuss the plausibility
of having dust in the IGM in the first place, \ie, the survival of dust
grains in IGM.

\subsection{Survival of dust particles}

Before the discussion of the heating process due to dust,
let us first consider the question of the survival of dust grains in
high-redshift Lyman-$\alpha$ clouds. As already mentioned, these clouds
have typical densities $\la 10^{-4} \, \rm cm^{-3}$ at high redshift,  
metallicities $\simeq
10^{-2.5} \,  Z_{\odot}$,
 and they are exposed to background radiation $\simeq 10^{-21}$
erg cm$^{-2}$ s$^{-1}$ sr$^{-1}$.
It is not clear  if these  clouds were enriched by metals driven out
of earlier generation dwarf galaxies  with galactic
winds or these metals come from {\it in situ \/} star formation; 
though the
small scatter in observed metallicities as $N_{\rm \scriptscriptstyle HI}$
changes by over four orders seems to favour the former explanation
(Cowie \& Songaila  1998).

In the model of enriching the IGM by galactic winds at high
redshifts (Silk, Wyse \& Shields 1987; Miralda-Escud\'e \& Rees 1997;
Nath \& Trentham 1997),
the enriched gas is swept along with the ambient IGM  by winds
erupting out of dwarf galaxies due to the injection
of energy by supernovae from an early bout of star formation.
Dust grains could form either
the stars or in the shells of supernovae (Dwek 1998). 
Inside the parent galaxies,
the processes of destruction of dust grains could be as efficient as
in the Galactic ISM. The destruction of dust grains in our Galaxy is
still not well understood (see McKee {\it et al. \/} 1987, and 
references therein).
However, detailed modelling show that
it is reasonable to assume that a   fraction ($\sim
10 \%$) of refractory dust survives the destruction process 
in ISM (Dwek 1998).

When the enriched material, presumably along with the dust grains, are
ejected in the IGM, the destruction processes become less important.
For sputtering, for example, the grain life time (above the
sputtering threshold) is,
\begin{equation}
t_{sp} \sim {N_a \over Y_{sp}} {1 \over \langle v \rangle n_s \sigma_g
Q_{coll}}
\sim 6 \times 10^{13} {1 \over (Y_{sp}/0.01) \, n_s T_4^{1/2}} \> {\rm s}
\>.
\end{equation}
Here, $N_a$ is the number of heavy atoms in the grain, $Y_{sp} \le 0.01$
is
the sputtering yield (number of atoms ejected from the grain per impinging
atom) (Draine \& Salpeter 1979),
$<v>$ is the thermal speed of the ambient particles with density $n_s$
and temperature $T=T_4 \, 10^4$ K, $\sigma_g=\pi a^2$ is the geometrical
cross-section of the grains with radius $a$ and $Q_{coll}$ is a measure
of the efficiency of collision. The numerical factor in the second
expressions assumes $N_a=3 \times 10^8$ and $a=1 \mu$ m.
In this equation, $n_s$ is the density in the shell of the radiative
shock ploughing through the IGM. Compare the above time-scale with the
cooling
time behind the shock (Shull \& Silk 1981),
\begin{equation}
t_c\sim 6\times 10^{12} {T_4^{1/2}\over (1+8.3\times 10^3 \zeta
T_4^{-1/2})n_s}{\rm ~s}\> ,
\end{equation}
where $\zeta=Z/Z_\odot$. Therefore, dust can survive sputtering in the
hot postshock region before it cools.

The Lyman-$\alpha$ clouds are exposed to UV background radiation which can
potentially destroy the dust particles from sublimation. 
However, the intensity
of background radiation is several orders below the ambient 
radiation in local
HII regions and it can be shown that  the grain temperature
is   close to that of the microwave
background radiation (Loeb \& Haiman 1997; Ferrara {\it et al. \/} 1998),
which is too low for sublimation to be important.

If Lyman-$\alpha$ clouds
are assumed to be star-forming regions then the  UV flux 
inside these clouds
can greatly exceed the background value (for dust heating
we do not consider this case). From deep, serendipitous
searches for redshifted Lyman-$\alpha$ and $H\alpha$ emission lines,
 it is possible to put an upper limit on the ionizing
intensity  (or equivalently on the average star-formation rate)
inside the clouds
(Hogan \& Weymann 1987). These
 searches give upper limits which  suggest that the Hydrogen-ionizing flux
inside the clouds do not exceed the background value by
 more than a factor of 10 (Thomson and Djorgovski 1995, 
Thomson {\it et al. \/}
1996). As the dust grain temperature, $T_d \propto J_\nu ^{1/5}$,
this additional UV flux does not appreciably enhance  it.

In the local ISM, other destruction mechanisms are cosmic ray erosion and
shattering from dust collisions .  The properties of cosmic
ray at high redshifts are largely unknown. However, even if we assume the
cosmic ray density in the clouds to be comparable to local ISM, the timescales
of dust erosion are $\ge 10^{10} \, \rm yrs$ (Draine \& Salpeter 1979).
The timescale of dust-dust collision in shocks in local ISM is
 $\simeq 10^{8} \, \rm yrs$ (Jones {\it et al. \/} 1996).
 In Lyman-$\alpha$ clouds
this time scale would be larger by a factor of the square of the
 ratio of ISM to
the cloud number density of grains. As this factor is $\ge 10^{7}$, 
dust-dust
collision can be neglected as an important destruction mechanism. 
Also, since the magnetic field
in the IGM is thought be small ($B \le 10^{-9}$ G; Kronberg 1994), other
destruction processes like non-thermal sputtering from betatron
acceleration
also become unimportant.

Therefore, we assume the dust to gas ratio to be proportional to that in
Galactic ISM with a factor $\zeta = Z/Z_{\odot}$.



\subsection{$\lyal$ absorbers}

It is clear from the Figures 5, 6 and 7 that dust heating can play
an important role in determining the physical state of the gas in $\lyal$
absorbers, if they can be considered as overdense regions in the IGM.
As mentioned in \S 3, the gas in $\lyal$ absorbers are most likely 
to behave somewhere between the two extreme cases considered here,
expand in two dimensions while being confined in one direction. In this
case the gas would behave close to the evolution in Model I, and dust
can heat up the gas instead of letting it cool.

The gas temperature in $\lyal$ absorbers, however, is not directly
observable. What is instead observed is the line width (the $b$
value), and lines can be both thermally and velocity broadened. 
However, Haehnelt, Rauch \& Steinmetz (1996) have argued that the
minimum value of $b$ in its distribution corresponds to the gas
temperature, at least for lines with $\nh \ga 10^{14}$ cm$^{-2}$.
Numerical simulations seem to show that lines with lower column
density are mostly velocity broadened (Miralda-Escud\'e \& Rees 1996).
In this context, it is interesting to note the recent observations
of Kim \etal (1997). They used Keck spectra of 5 quasars at different
redshifts and found that the minimum value of $b$ increased in
time. We show the corresponding values of temperature in Figure 7 as
thick lines. Although it is difficult to draw any firm conclusion,
the evolution of temperature does suggest that dust heating can
contribute towards raising the gas temperature in $\lyal$ absorbers
as required by these observations.
Note, that an alternative 
explanation for increase of the minimal Doppler width with decreasing 
$z$ has been suggested recently by Giallongo (1997) as a result of 
decrease of the softness parameter in framework of the photoionized 
heating model (Ferrara \& Giallongo, 1996). 

A note on the uncertainties in our calculations due to our assumptions
concerning the background radiation is in order here.
Unlike simple photoionization heating, heating due to dust grains
is proportional to the intensity of the background radiation. This
makes our calculations sensitive to the uncertainties in the background
radiation, which include uncertainties in the QSO luminosity function,
the contribution from young galaxies and the opacity due to $\lyal$
systems. As an example, using the LA model of absorption from $\lyal$
clouds of Meiksin \& Madau (1993) (see, Sethi \& Nath 1997), instead
of model A3 of Giroux \& Shapiro (1996), would increase the background
flux and the dust heating rate.

Since dust heating rate is proportional to the flux of photons, we
would expect a rise in temperature in absorbing systems close to
quasars. The uncertainties in various parameter, however, may make
this prediction difficult to test. Another prediction of our
calculation is that, with a large scatter in the metallicity
and also in the value of $z_{in}$, there would be a scatter in the
temperature and in the minimum width in the distribution of $b$
at a given redshift.

As we have shown in \S 3, if there is a low density `intercloud' part
of the IGM, then the thermal evolution of this component of IGM would
be much different with the inclusion of dust heating than if only
photoionization heating were used. This heating would also depend on
the metallicity, and the metallicity of the IGM is not expected to be
homogeneous. Nath and Trentham (1997) have estimated the probability
distribution function of metallicity in the IGM, in the case of IGM
enrichment from galactic winds, and found the standard deviation of
the metallicity of order $0.01 Z_{\odot}$. Rauch \etal (1997) found
from the comparison of data with their numerical simulation a scatter
of an order of magnitude in the metallicity of $\lyal$ absorbers. It,
therefore, is reasonable to assume that there are regions in the IGM
with higher dust content than we have considered above, and so,
hotter in general. (Of course, the actual temperature would depend
on the thermal history of the region.) 

Finally, we estimate the metallicity that is needed for dust heating
to be more important than photoionization of a primordial gas. From
fig. 6, we estimate this lower limit for the accepted QSO
type spectrum to be $Z/Z_{\odot} \ga 10^{-2.5}
\, (n/10^{-5} \, cm^{-3})
\, (J_{912,-21}/0.5)^{-0.7}$, when the grain mixture is dominated
by silicates (as in LMC or SMC type of mixture).

\subsection{Feedback from IGM with dust heating}

The feedback of the intergalactic medium on the process of galaxy formation
is an interesting phenomenon and has been discussed in by several authors.
The reionization and reheating of the intergalactic medium inhibit the
formation of small galaxies by the suppression of cooling and because of the
high temperature of the IGM gas (Efstathiou 1992, Thoul \& Weinberg 1996).
Here we briefly discuss the implications
of dust photo-electric heating for such feedback mechanisms.

First, unlike photoionization heating of a primordial gas, the dust
photo-electric heating is proportional to the UV background radiation
intensity. Fluctuations in the intensity and spectrum of the background
radiation would result in variation in the inhibiting efficiency. However,
even for silicates and for QSO type spectrum, the gas temperature is
expected to change only by a factor
$\sim 1.5$, thereby increasing the Jeans mass by a factor of
$\sim 2$. The dependence of the dust heating on the dust-to-gas ratio, however,
has more interesting effects.
When dust
heating dominates in gravitationally confined (i.e. with constant or
growing density) clouds, and the cooling is determined mainly by
bremsstrahlung emission, $\Lambda(T)\propto T^{1/2}$, the gas
temperature can be estimated
from Eq.~\ref{eq:quahet} as

\begin{equation}
T\propto {\cal D}^{(\alpha+\beta+1)/(\alpha+\beta)}J^{2/(\alpha+\beta)}
n^{-2/(\alpha+\beta)},\>
\end{equation}
where ${\cal D}$, is the dust-to-gas ratio.
Therefore, the enrichment of the gas (higher ${\cal D}$)
results in the increase of gas temperature (and Jeans mass,
$M_J\propto {\cal D}^{3(\alpha+\beta+1)/2(\alpha+\beta)}$), and thus
suppresses subsequent formation of low mass galaxies. One
can speculate from these simplified arguments that dust
heating tends to suppress the progressive enrichment of the IGM with heavy
elements. However, if the metallicity of enriched IGM gas exceeds a  critical
value $\zeta>\zeta_c$, when radiative cooling from heavy
elements dominates over bremsstrahlung cooling, metal-enriched regions cool
efficiently and dust heating does not work anymore as a negative
feedback. For cooling due to C~IV, which is the most dominant heavy
element for cooling in these conditions, we estimate that $\zeta_c \sim 0.05 \,
(n/10^{-4})^{-1}$, for $J_{912,-21}=1$, $\alpha=1$, and $T=5 \times 10^4$ K.
Note also, that dust heating can work as a positive feedback
(stimulating formation of regions of low temperature due to thermal
instability) if hard spectrum
($\alpha\leq 0.5$) and large ($a \geq 0.01~\mu$m) silicate grains
dominate, though this case seems to be unrealistic.

\section{Conclusion}

We summarise our findings:

(a) It is reasonable to associate dust grains with the metallicity
observed in high redshift $\lyal$ absorbing systems, as the probability
of the survival of dust grains during the process of spreading the metals
and dust in the IGM and in the $\lyal$ systems is high.

(b) With the presence of the UV background radiation at high redshift,
the dust particles in the IGM acquire charges which are much larger 
than is typical
in the ISM, especially because of the low density of gas and high
intensity of hard UV and X-ray photons. We illustrate
this effect using two models of the UV background radiation, described in
Table 1. Figures 1 and 2
show the dependences of the grain potential on gas density and grain size for
graphites and silicates. Silicates in general have larger charges, and harder
spectra help to produce larger grain potentials.

(c) This results in a higher mean energy of the photo-ejected electrons, and
therefore, yields an important source of heating of the IGM gas. We show the
volume heating rate as functions of the gas density, grain size distribution
and grain type. Silicates fare better as heating agents, and smaller
grains dominates the heating rate in general. We estimate that
dust heating would be important for our QSO type spectra for a metallicity
$Z/Z_{\odot} \ga 10^{-2.5} \, (n/10^{-5} \, cm^{-3}) \, J_{912,-21}^{-0.7}$.

(d) Using these heating rates, and grain mixtures of the LMC type, we compute
the temperature evolution of the gas in overdense regions of the IGM, with
overdensity of $\sim 10$, for various initial redshifts, initial temperatures
and metallicities. We assume that the $\lyal$ systems evolve in a passive
manner, driven by the evolution of the background UV radiation and the 
boundary condition of the cloud expansion.
We show the evolution for overdense regions which (i) 
adiabatically expand and keep the overdensity constant and (ii) which are
confined either by shocks or gravity and keep the (proper) density constant.
The difference in the temperatures in the cases when dust heating is included
and when it is not, is of order of a factor of $\sim 2$ for our hard
QSO type spectra, for a metallicity
of $Z/Z_{\odot}=0.01$ in a Hubble time at $z\sim 3$ for $\Omega_b=0.03$.

(e) We have speculated on the connection between the dust heating in such
systems and the observed increase in the minimum width of $\lyal$ absorption
lines in time (most probably indicative of increase in the gas temperature).
We have also discussed the possible increase in the gas temperature of $\lyal$
absorption systems near QSOs due to increased UV flux, and the patchiness in
the temperature of the IGM due to (i) density inhomogeneity and the (ii) 
inhomogeneity in the metallicity.

\medskip
\noindent
{\bf Acknowledgements : }
We thank A. Ferrara and E. E. Salpeter for valuable discussions and
the anonymous referee for his comments. 
YS acknowledges the hospitality of IUCAA, India, where this work began, 
the hospitality of Astronomishes Institut, Ruhr
Universit\"at Bochum at final stages of the work.
and financial support from Italian Consiglio Nazionale delle
Ricerche within NATO Guest Fellowship programme 1996 (Ann. No. 219.29).

\bs

\clearpage

\centerline{{\bf Figure captions}}
\bigskip
\noindent
Figure 1(a) --
Grain potential $eU/kT$ is plotted against ambient density $n_e$
for QSO and QSO+YG
spectra for graphites for two values of $J_{912,-21}=0.5, 0.05$
and for $a=10^{-5}$ cm.

\bigskip
\noindent
Figure 1(b) --
Same as in Fig.1a but for silicates.

\bigskip
\noindent
Figure 2(a) --
The grain potential is plotted against $a$ for graphites
for two incident photon spectra (QSO and QSO+YG) for $J_{912,-21}
=0.5, 0.05$, and for $n_e=10^{-5}, 10^{-4}$ cm$^{-3}$.

\bigskip
\noindent
Figure 2(b) --
Same as in Fig. 2a but for silicates.

\bigskip
\noindent
Figure 3 --
Mean energy of photo-emitted electrons is shown against ambient density for
$a=10^{-5}$ cm, for QSO and QSO+YG spectra, for graphites.

\bigskip
\noindent
Figure 4 --
Same as in Fig. 3 but for silicates.

\bigskip
\noindent
Figure 5 --
The mean electron energy $E(a)$ is plotted against $eU/kT$ for
various sizes, photon spectra and grain types, at $T=10^4$ K.
The solid lines are for silicates and the dotted lines are for
graphites. The thicker lines refer to QSO and thin lines denote
QSO+YG type spectra. The upper set of curves are for $a=10^{-5}$cm
and the lower set for $a=10^{-6}$ cm.

\bigskip
\noindent
Figure 6 --
Integrated (over size distribution) heating rate
for graphites and silicates vs ambient density
for QSO (solid lines) and QSO+YG (dotted lines) spectra. The upper set of
curves are for $J_{912,-21}=0.5$ and the lower set for 
$J_{912,-21}=0.05$. Photoionization heating rate for a primordial
gas is shown for QSO (short dashed lines) and QSO+YG (long dashed
lines). 

\bigskip
\noindent
Figure 7 --
The heating rate is shown as a function of the lower limit of the grain
size (a), for silicates (solid lines) and graphites (dotted lines) for
two different densities $n=10^{-4}, 10^{-5}$ cm$^{-3}$, for QSO spectrum
and at $T=10^4$ K.

\bigskip
\noindent
Figure 8(a) --
Temperature evolution for clouds with constant density (left panels)
and constant overdensity (right panels) for graphites.
The evolution is shown for $z_{in}=3$ \& $4$.
The top and bottom panels are for QSO and QSO+YG spectra respectively.
The dotted lines show the case with only photoionization heating.
the solid lines correspond to the case where dust heating is included
for a metallicity of $Z/Z_{\odot}=0.01$, for $\Omega_b=0.03$, and
$T_{in}=10^4$ K. Thermal history with the QSO spectrum is added also 
with $T_{in}=2\times 10^4$ K. 

\bigskip
\noindent
Figure 8(b) --
The same as Fig8a but for silicates.

\bigskip
\noindent
Figure 9 --
Temperature evolution for clouds with constant density (left panel) 
and constant overdensity (right panel) for QSO spectrum for a mixture
of grains of LMC type. The evolution is shown for $z_{in}=3$ \& $4$.
The dotted, dashed and solid lines correspond
to the cases with only photoionization, with dust with metallicities
$Z/Z_{\odot}=0.003$ and $0.01$. The three sets of curves correspond
to initial temperatures $T_{in}=1,2,5 \times 10^4$ K.
The thick solid lines show the temperatures  corresponding to the
minimum $b$ values from Kim \etal (1997).

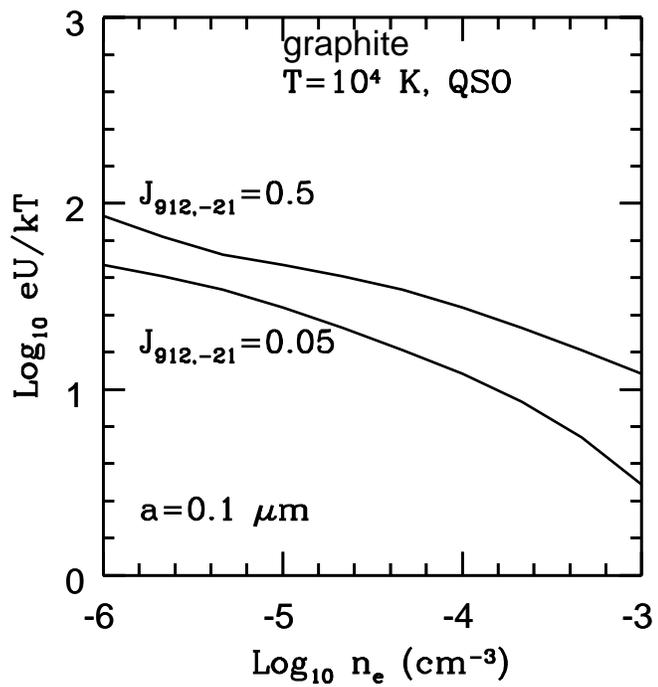
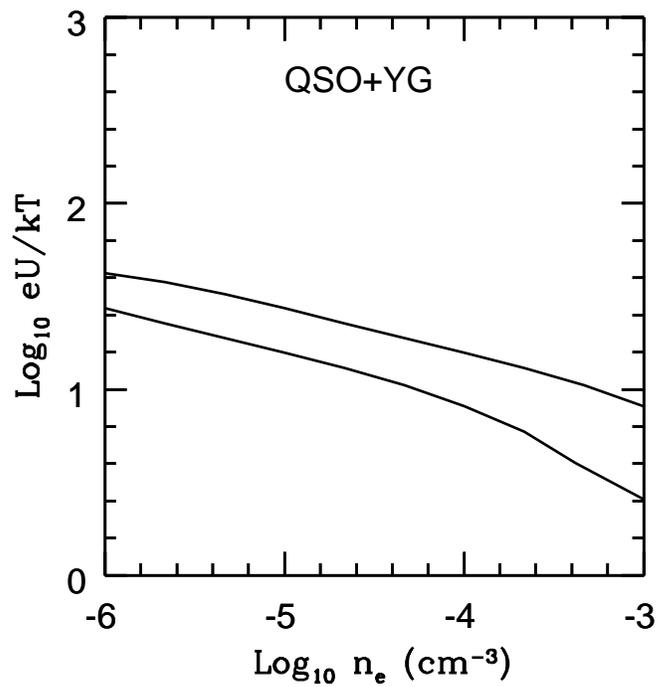
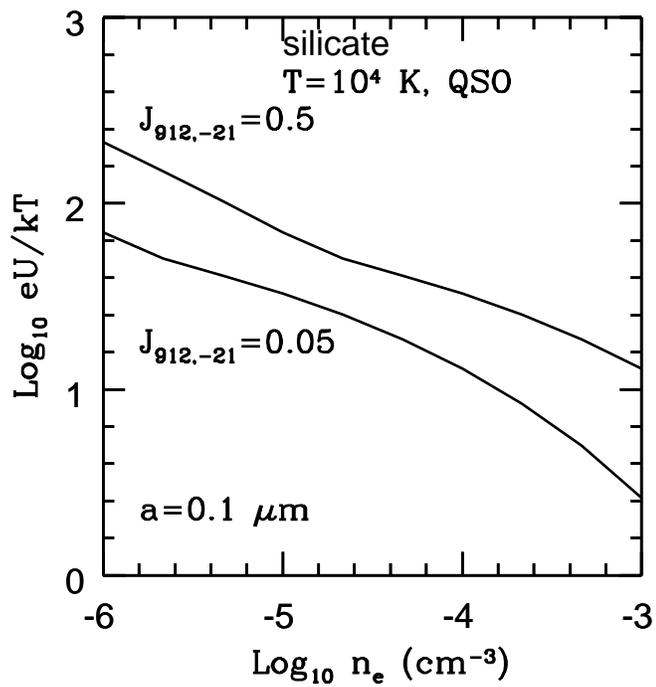
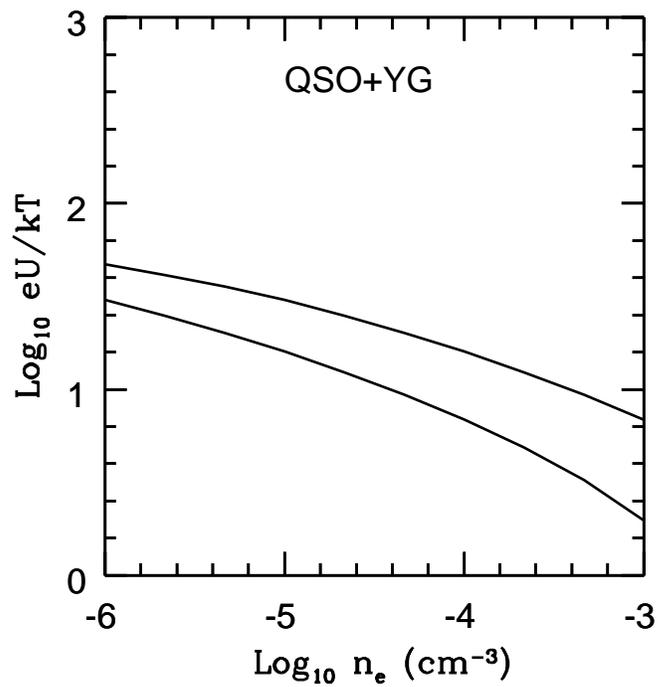
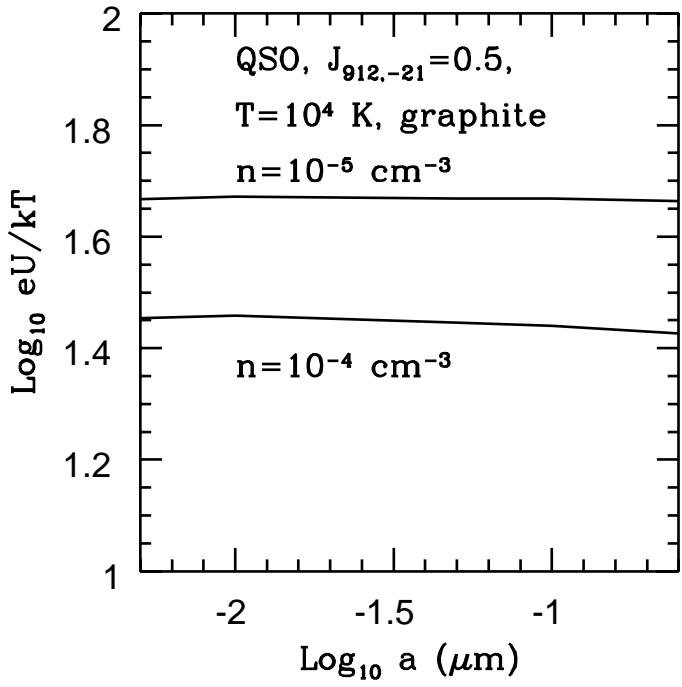
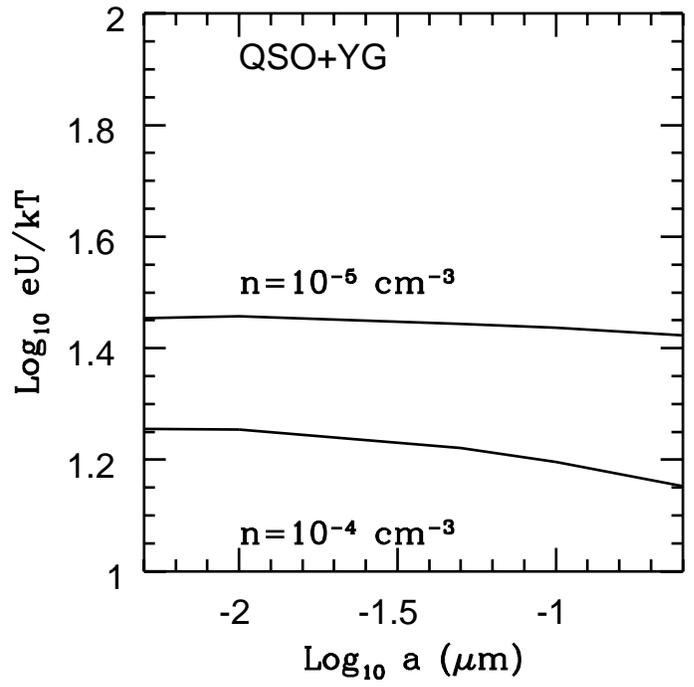
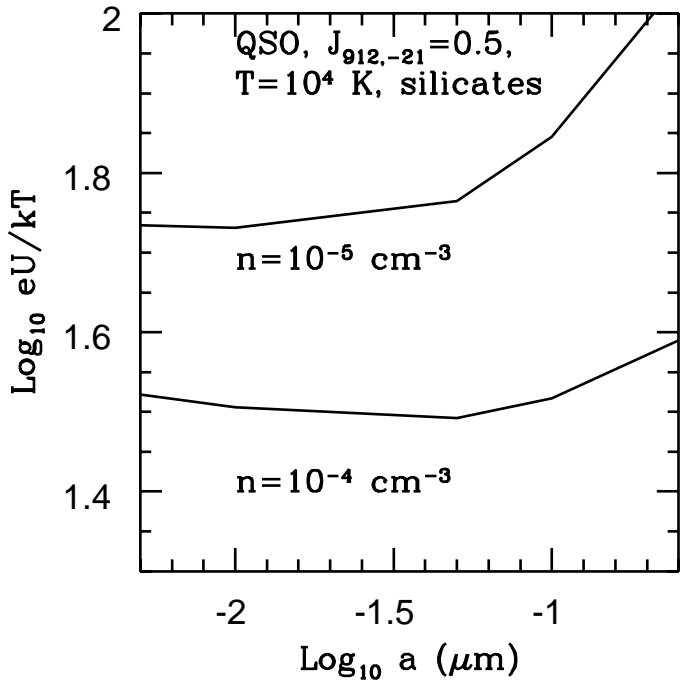
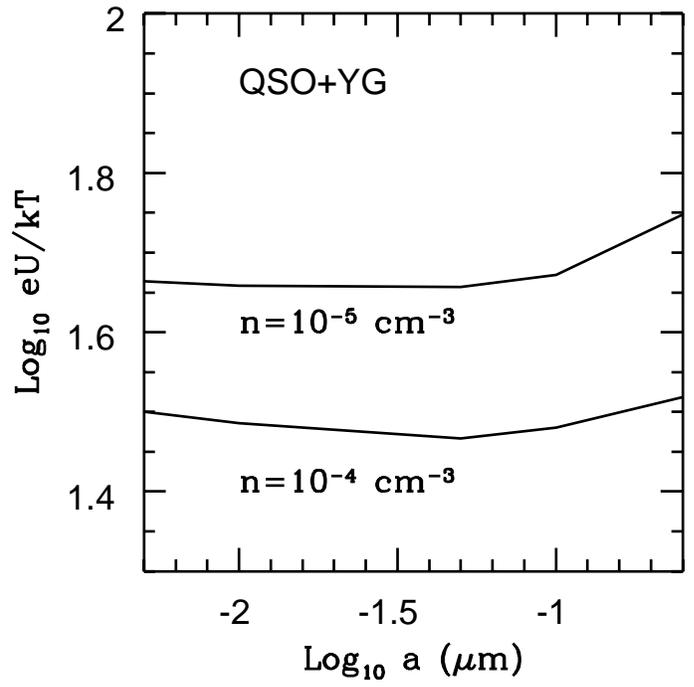
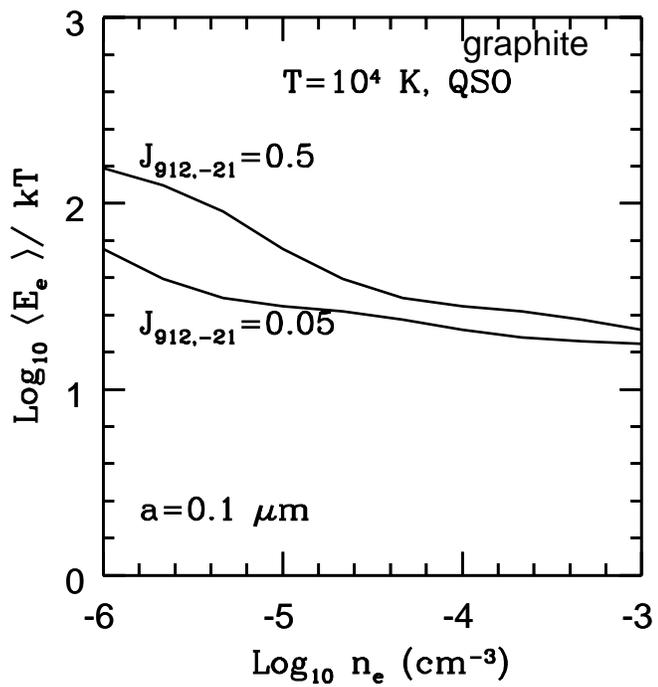
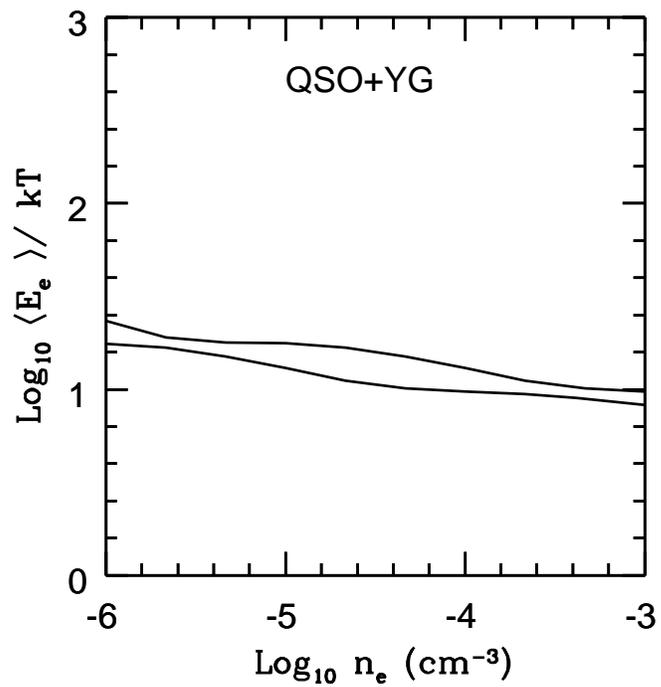
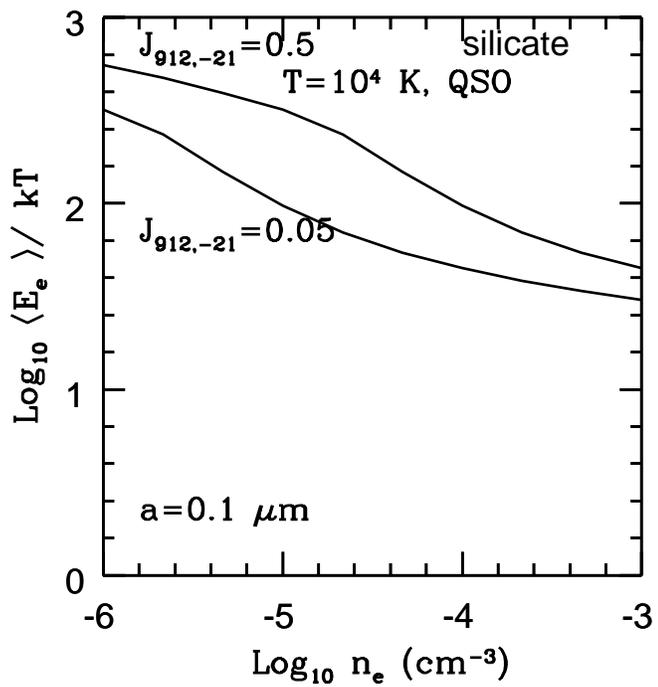
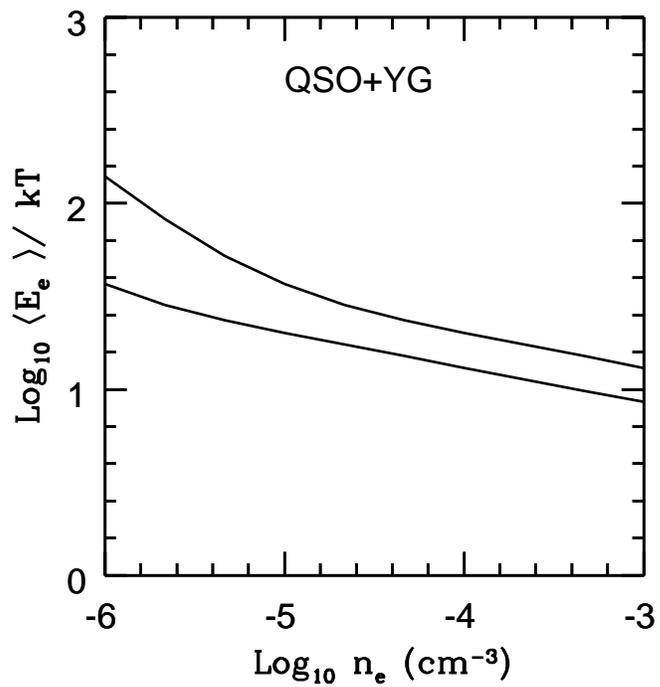
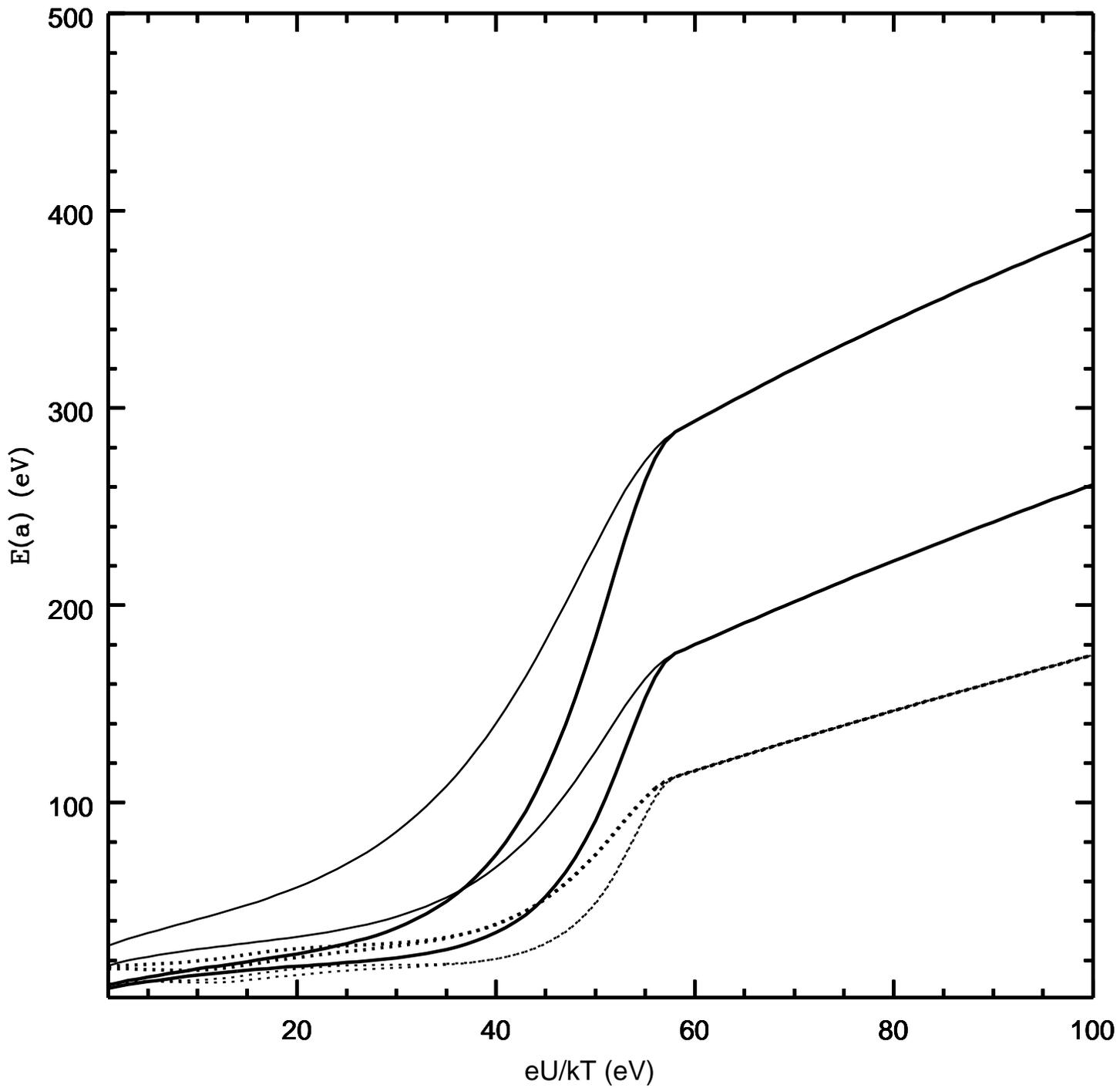
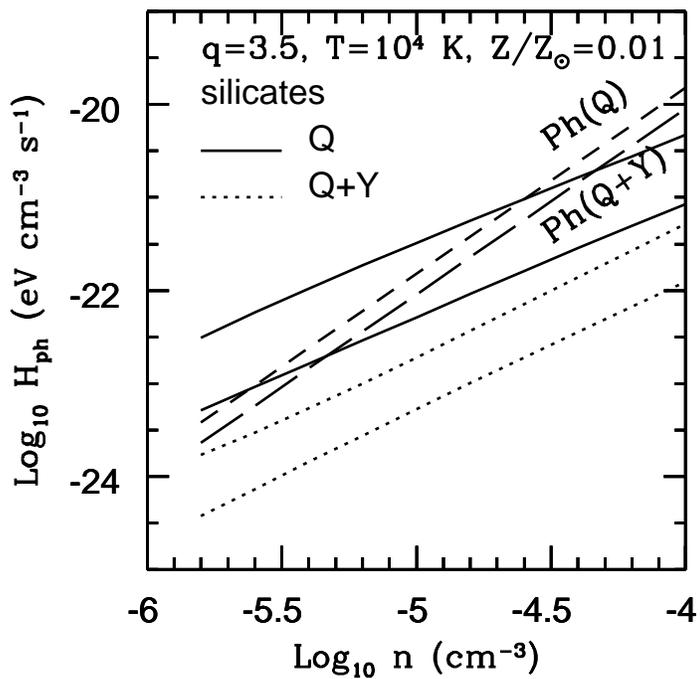
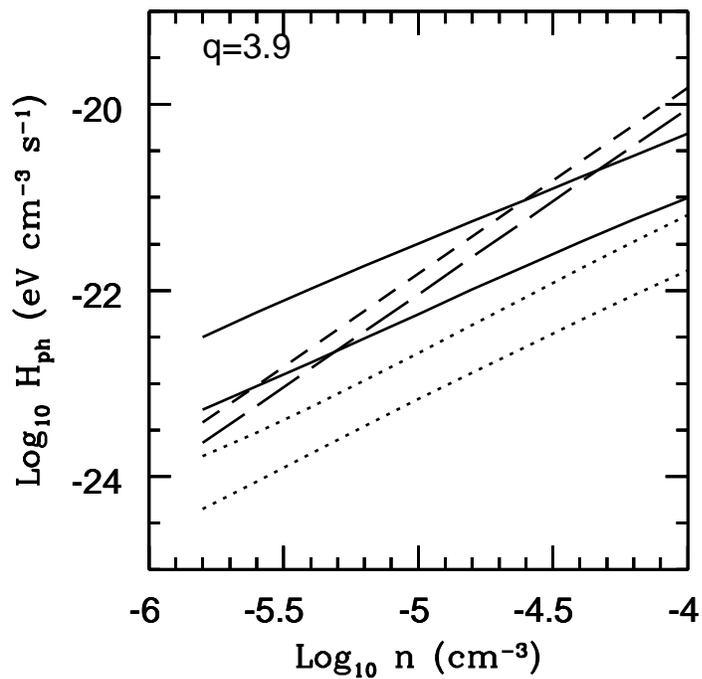
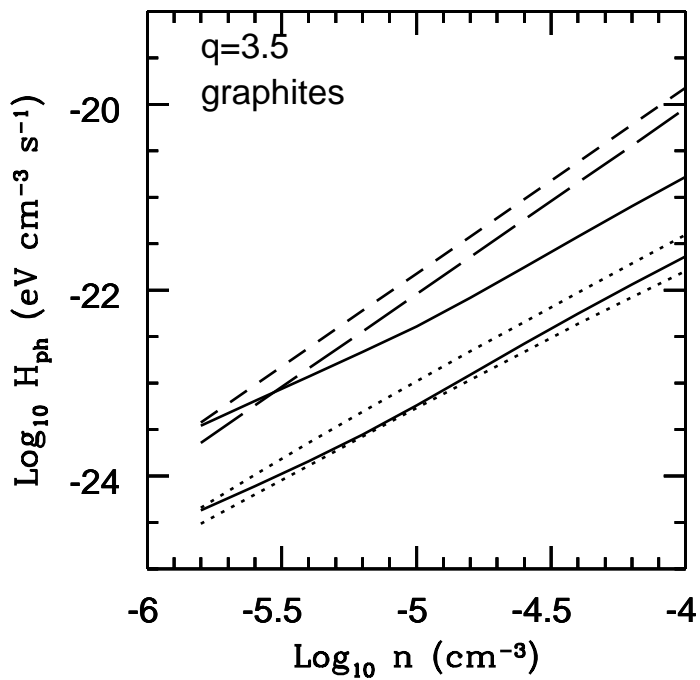
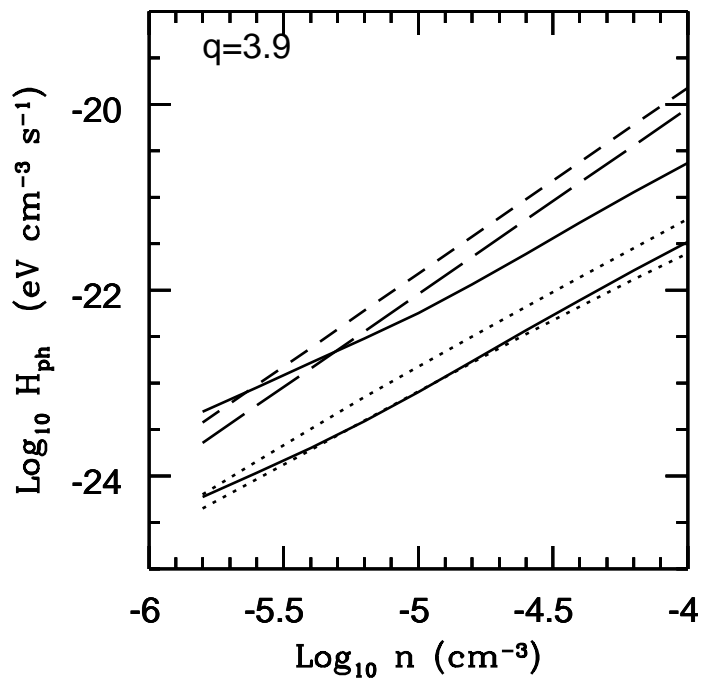
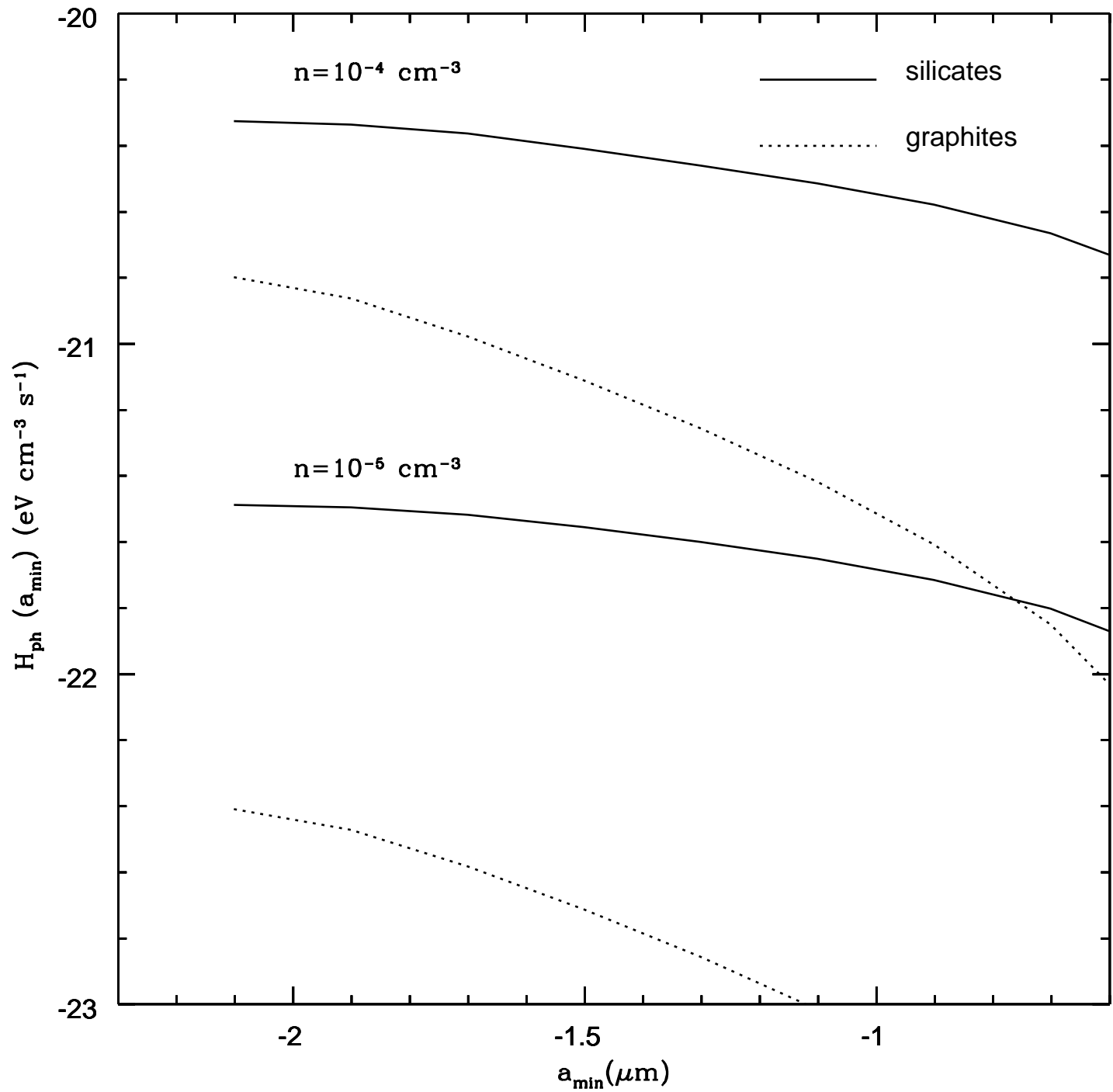
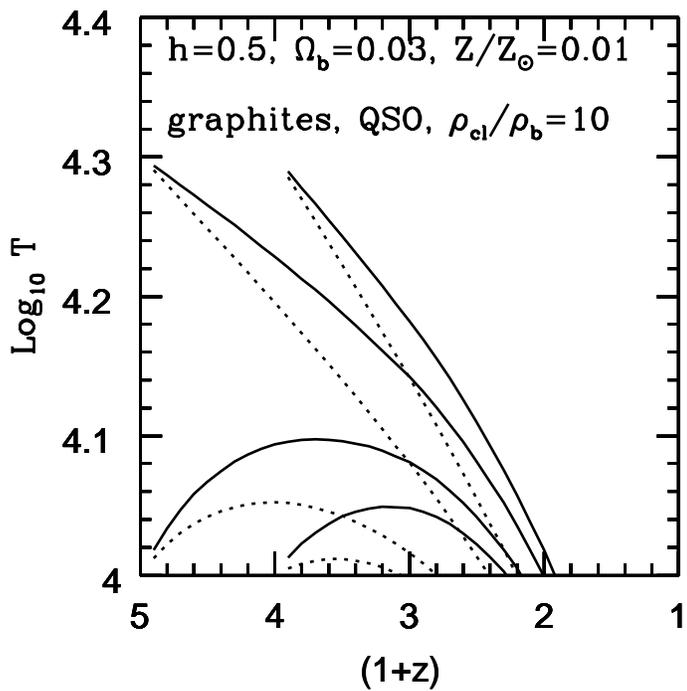
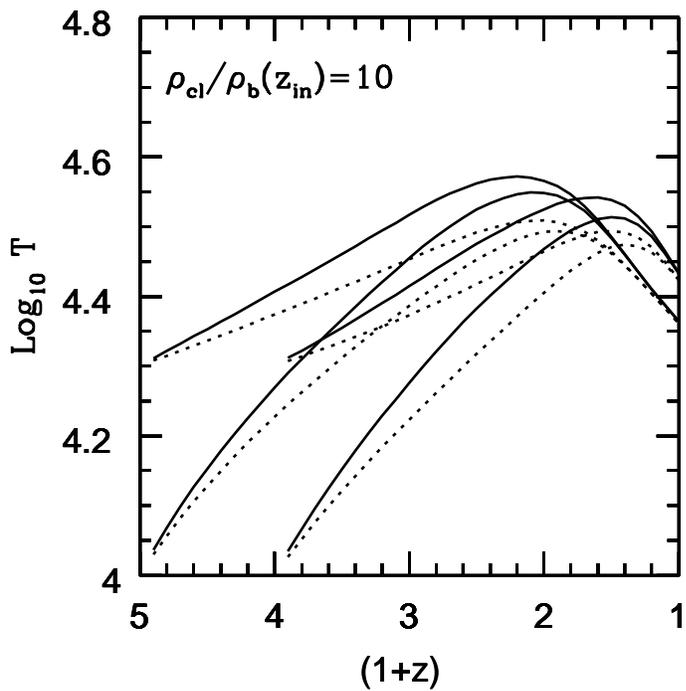
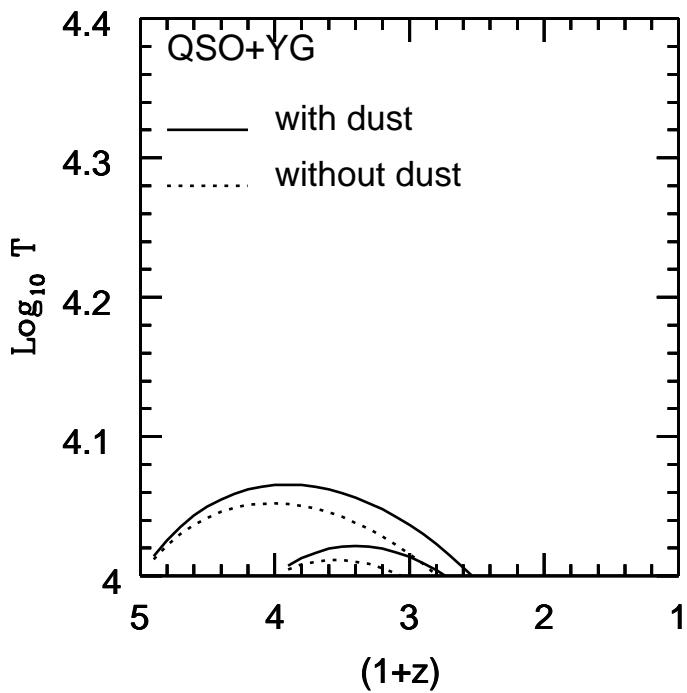
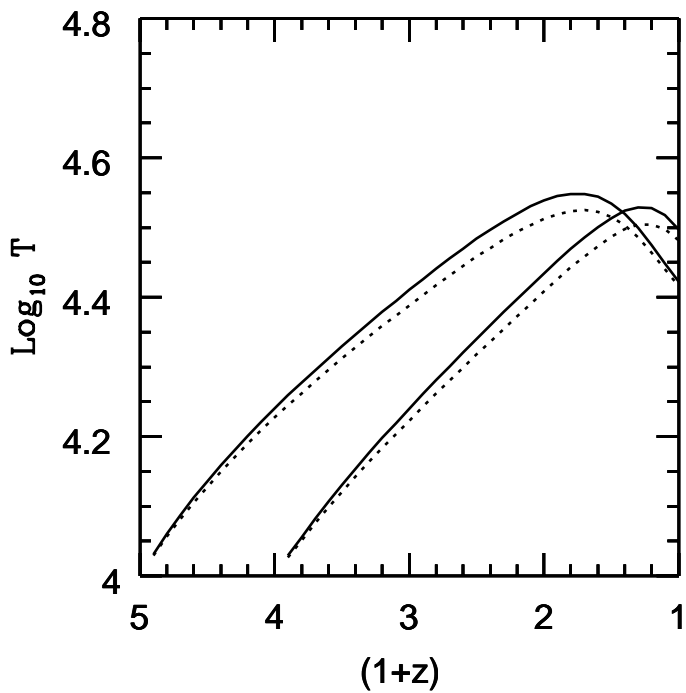
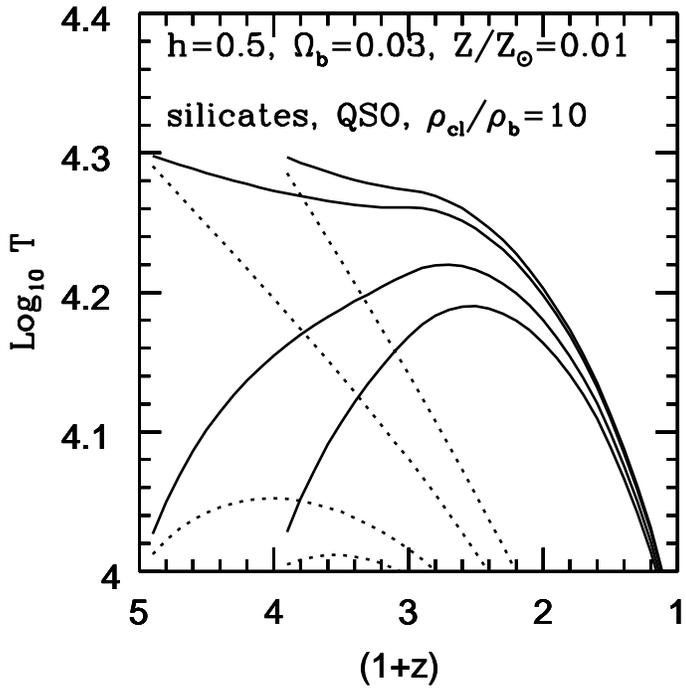
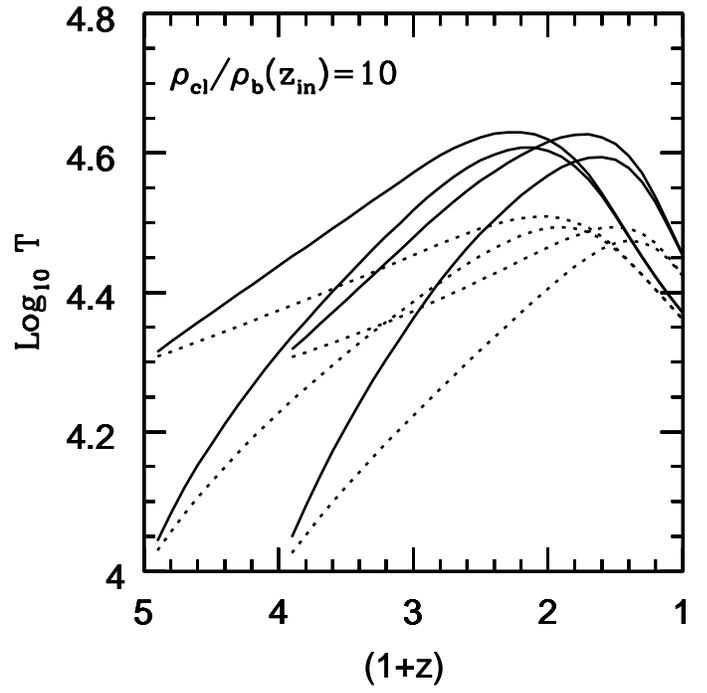
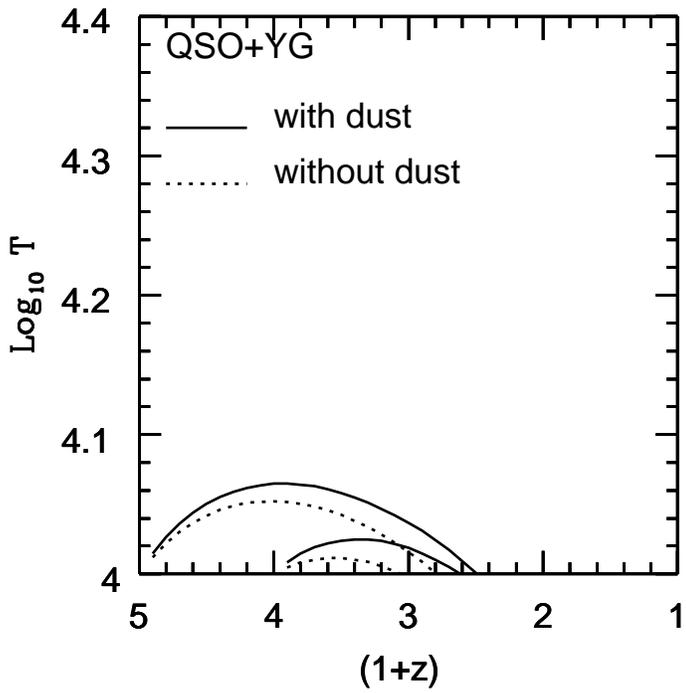
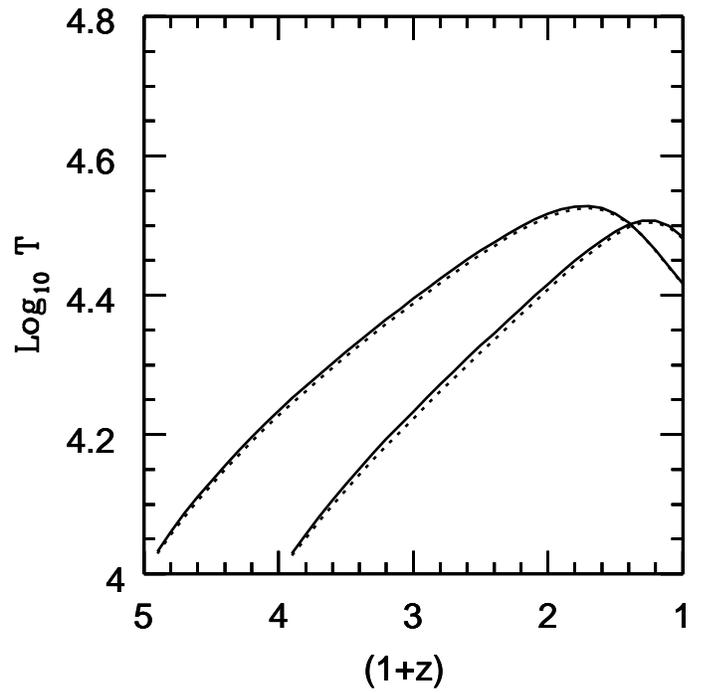
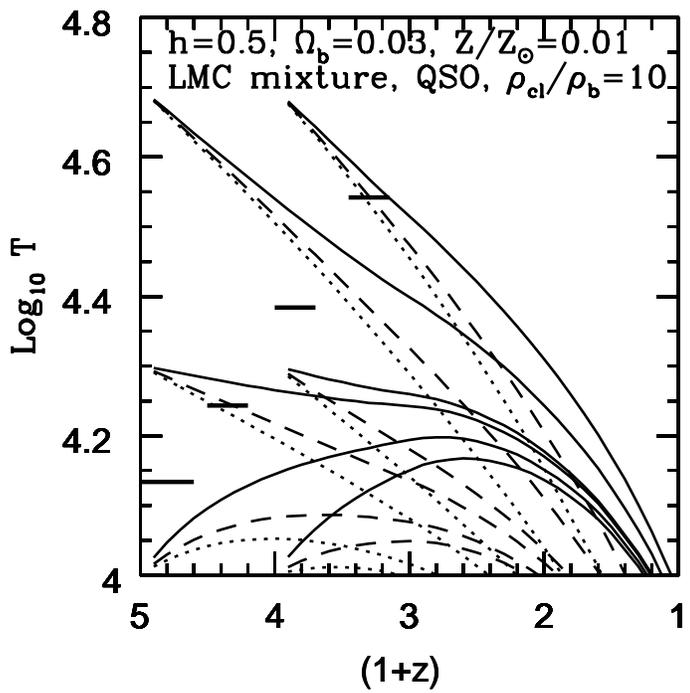
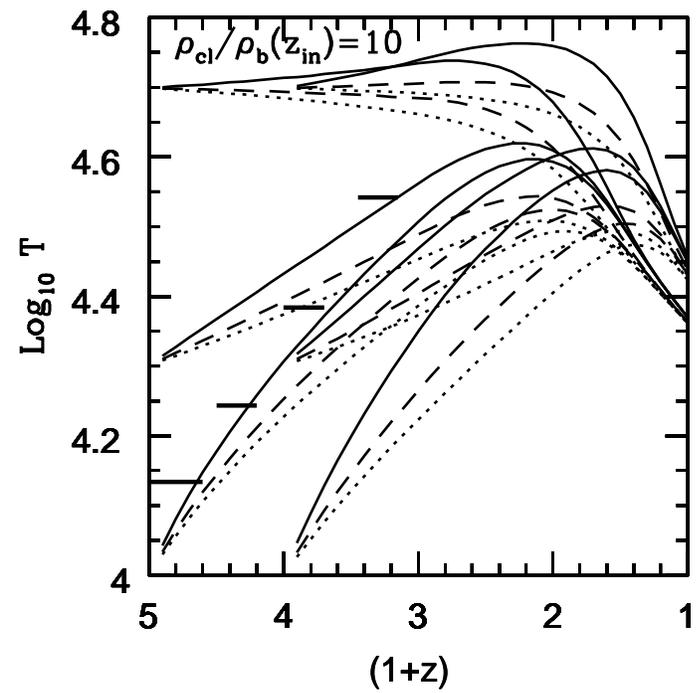
\end{document}